\documentclass[]{jkas} 
\usepackage{booktabs} 
\usepackage{makecell}
\usepackage{multirow} 
\usepackage{tikz}
\usetikzlibrary{fit}
\usetikzlibrary{positioning}
\usepackage{setspace}

\def\year{2024} 
\def\volume{56} 
\def\issue{1} 
\def\beginpage{1} 
\def\received{---} 
\def\accepted{---} 
\def\published{---} 
\date{Received \received; Accepted \accepted; Published \published}

\title{%
KMTNet Synoptic Survey of Southern Sky II: Data Reduction and Real-Time Transient Detection Pipeline
}

\author[1]{Mankeun Jeong}{0009-0003-1280-0099}
\author[1$\star$]{Myungshin Im}{0000-0002-8537-6714}
\author[2]{Joonho Kim}{0000-0002-1294-168X}
\author[1]{Seo-Won Chang}{0000-0002-3118-8275}
\author[1]{Sungho Jung}{0009-0009-5870-4266}
\author[3]{Chung-Uk Lee}{0000-0003-0043-3925}
\author[3]{Dong-Jin Kim}{0000-0002-4292-9649}
\author[1]{Bomi Park}{0009-0008-3499-3043}
\author[1]{Jaewon Lee}{0009-0005-3910-0337}
\author[1]{Jiseop Shin}{0009-0005-3944-1457}
\author[1]{Changwan Kim}{0000-0002-5266-1658}
\author[1,4]{Gregory S. H. Paek}{0000-0002-6639-6533}

\affil[1]{SNU Astronomy Research Center, Astronomy Program, Department of Physics and Astronomy, Seoul National University, Gwanak-gu, Seoul 151-742, Republic of Korea}
\affil[2]{Daegu National Science Museum, 20, Techno-daero 6-gil, Yuga-myeon, Dalseong-gun, Daegu 43023, Republic of Korea}
\affil[3]{Korea Astronomy and Space Science Institute, Daejeon 34055, Republic of Korea}
\affil[4]{Institute for Astronomy, University of Hawaii, 2680 Woodlawn Drive, Honolulu, HI 96822, USA}

\def\corrauthor{%
M. Im, \email{myungshin.im@gmail.com}
}

\def\runningauthor{%
Jeong et al.
}

\def\runningtitle{%
KS4 DR1 Reference Image Pipeline
}

\def\keywords{%
surveys --- astrometry --- techniques: photometric --- techniques: image processing --- methods: data analysis --- gravitational waves}

\def\abstracttext{%
We present a comprehensive pipeline developed for the image processing of the KMTNet Synoptic Survey of the Southern Sky (KS4) Data Release 1. This pipeline encompasses several key processes, including data quality assurance, astrometry, photometric zero-point (ZP) calibration, bad pixel masking, image stacking, and difference image analysis (DIA). The astrometric solutions were validated by cross-matching with the Gaia EDR3 catalog, achieving sub-pixel astrometric accuracy ($<$0.4 arcsec). To ensure spatial consistency, we divided each image into multiple subsections and confirmed that astrometric accuracy was maintained even at the edges. We performed a two-stage photometric calibration. Initial ZP solutions were computed for each individual image frame using the APASS DR9 and SkyMapper DR3 catalogs. Subsequently, we corrected residual spatial variations in the stacked images using Gaia XP photometry. This procedure yielded a 5$\sigma$ depth of 22–23~AB mag across the $BVRI$ bands, with root-mean-square errors of approximately 0.03 mag when referenced to Gaia stars in the magnitude range of 14--19 mag. The processed KS4 images span over 4,000 deg$^2$ of the southern sky, providing reference images suitable for DIA. This publicly available pipeline also supports real-time processing of newly acquired images, enabling prompt transient detection. We demonstrate its effectiveness through successful applications in gravitational-wave follow-up observations.
}

\begin{document}
\jkashead 


\section{Introduction}\label{sec:1.Introduction}

In the era of time-domain astronomy, the Korea Microlensing Telescope Network (KMTNet) offers significant advantages through its continuous, wide-field observing capability. KMTNet consists of three identical 1.6-meter telescopes, each equipped with a camera providing a 2~$\times$~2 square degree field of view (FOV). These telescopes are installed at the Cerro Tololo Inter-American Observatory (CTIO), the South African Astronomical Observatory (SAAO), and the Siding Spring Observatory (SSO) \citep{2016JKAS...49...37K}. This global distribution enables near-continuous sky coverage with minimal interruption. Such capabilities are ideally suited for the discovery and monitoring of transient phenomena, as demonstrated by several successful target-of-opportunity (ToO) and monitoring programs \citep{2016SPIE.9906E..4IM, 2018JKAS...51...89K, 2019ATel12438....1S, 2024Natur.626..742Y}.

Furthermore, in the context of gravitational-wave (GW) astronomy, KMTNet’s prompt response across the southern hemisphere is particularly valuable. The discovery of GW170817, which was accompanied by a rapidly fading kilonova, required deep and fast-response optical imaging for successful identification \citep{2017ApJ...848L..12A, 2017ApJ...849L..16I, 2017Natur.551...71T}. Since that event, timely electromagnetic observations have been recognized as crucial for characterizing a wide range of GW sources and for constraining the nature of their progenitors and environments \citep{2019ApJ...885L..19C, 2019ApJ...880L...4H, 2020MNRAS.497..726G, 2020ApJ...905..145K}. KMTNet can reach a 5$\sigma$ limiting depth of $\gtrsim$22 mag with 480-second exposures, making its rapid and sensitive follow-up observations well suited for surveying significant cosmic volumes \citep{2021ApJ...916...47K, 2024ApJ...960..113P, 2025ApJ...981...38P, 2025arXiv250315422P}.

However, effective transient identification relies on the availability of pre-existing reference images. Building a large, deep, and uniform reference image database is therefore essential for real-time transient detection, especially given that GW localization area can span several hundred square degrees across unpredictable parts of the sky \citep{2020LRR....23....3A}. Despite this need, no public reference image database has previously been available for KMTNet.

The KMTNet Synoptic Survey of the Southern Sky (KS4) was initiated to resolve this limitation (Im et al., in preparation). This survey targets the southern sky at declinations between $-85^\circ$ and $-30^\circ$ (excluding most of the Galactic plane) with the aim of building a comprehensive reference image database. This paper presents the data reduction pipeline and the procedures used for the data in Data Release 1 (DR1) of KS4, covering observations from November 2019 to December 2023. The detailed survey footprint and field distribution of the $4,000 \text{deg}^2$ covered in DR1, as well as the anticipated future coverage, are illustrated in Figure 1 of the companion paper \citep{2026JKAS...Chang} and are available on the KS4 project page at NOIRLab’s Astro Data Lab\footnote{\url{https://datalab.noirlab.edu/data/ks4}}. This dataset includes a catalog of over 200 million sources (with signal-to-noise ratio above 5) and reference images covering more than 4,000 deg$^2$ across 979 KS4 fields, observed in the Johnson–Cousins $BVRI$ filters with at least four 120-second exposures per filter per field.

This data reduction pipeline includes data quality assessment, astrometric and photometric calibration, and image stacking. Previous efforts, such as the Deep Ecliptic Patrol of the Southern Sky \citep[DEEP-South;][]{2016IAUS..318..306M, 2018JKAS...51..129C} and the KMTNet Supernova Program \citep[KSP;][]{2016SPIE.9906E..4IM, 2017ApJ...848...19P, 2019ApJ...885...88P}, have demonstrated the scientific utility of KMTNet for transient science, providing calibrated photometry and reduction procedures for specific targets. Building on that foundation, our pipeline extends this capability to full-sky calibration across thousands of square degrees, incorporating advanced photometric calibration techniques using Gaia XP synthetic photometry \citep{2023A&A...674A...1G}. The final stacked images reach 5$\sigma$ depths of 22--23~AB mag in the $BVRI$ filters, with a photometric root-mean-square error (RMSE) of approximately 0.03 mag for sources brighter than 19~AB mag.

In addition to reference image construction, the KS4 pipeline is designed to handle real-time processing of newly acquired images from ToO observations using the same modular framework. This includes immediate reduction of ToO imaging data obtained during time-critical follow-up campaigns such as GW events. By applying consistent procedures—data quality assessment, astrometric and photometric calibration, and image subtraction—the pipeline ensures compatibility with pre-built reference images and minimizes latency in transient identification. Recent large-scale surveys such as SkyMapper \citep{2017PASA...34...30S}, the Dark Energy Survey \citep{2018PASP..130g4501M}, and Pan-STARRS \citep{2020ApJS..251....3M} have similarly demonstrated that combining deep reference images with automated pipelines is essential for maximizing time-domain science returns. Following this approach, we have made the KS4 DR1 reference image set and the reduction pipeline publicly available\footnote{\url{https://github.com/jmk5040/KMTNet_ToO}}, thereby extending these capabilities to the broader KMTNet community.

Lastly, we present selected transient follow-up cases associated with GW events detected by the Advanced Laser Interferometer Gravitational-Wave Observatory (aLIGO; \citealt{2015CQGra..32g4001L}), the Advanced Virgo detector (aVirgo; \citealt{2015CQGra..32b4001A}), and the Kamioka Gravitational Wave Detector (KAGRA; \citealt{2021PTEP.2021eA101A}), which collectively form the LIGO–Virgo–KAGRA (LVK) collaboration. These cases include the events S230518h \citep{2023GCN.33816....1L}, S240422ed \citep{2024GCN.36240....1L}, S240915b \citep{2024GCN.37513....1L}, S250206dm \citep{2025GCN.39231....1L}, and S250830bp \citep{2025GCN.41607....1L}. For each of these follow-ups, the KS4 reference images and reduction pipeline enabled efficient image subtraction and successful transient identification. These results highlight the practical value of KS4 in real-time ToO campaigns, reinforcing its scientific utility in the context of multi-messenger astronomy.

This paper is organized as follows. Section~\ref{sec:2.Data} outlines the KMTNet data structure and quality selection criteria. Section~\ref{sec:3.Data Calibration Methods} describes the reduction pipeline, including astrometric and photometric calibration, bad pixel masking, and stacking procedures. Calibration performance and practical applications in ToO observations are presented in Section~\ref{sec:4.Result}. The current limitations of the pipeline and potential directions for future development are discussed in Section~\ref{sec:5.Discussion}. Section~\ref{sec:6.conclusion} concludes with a summary of key results.


\section{Data} \label{sec:2.Data}

\subsection{KMTNet Image Data Structure} \label{sec:2.1.KMTNet Image Data Structure}

The KMTNet camera consists of a mosaic of four charge-coupled device (CCD) chips (each referred to as a ``chip image'') with dimensions of 9216~$\times$~9232 pixels and a pixel scale of 0.4~arcsec~pixel$^{-1}$. Each chip covers a FOV of approximately 1.024~deg~$\times$~1.026~deg, resulting in a total effective FOV of roughly 2~deg~$\times$~2~deg for the entire camera.

The image is aligned with equatorial coordinates such that the x-axis corresponds to lines of constant declination (with right ascension decreasing in the positive x-direction), and the y-axis corresponds to lines of constant right ascension (with declination increasing in the positive y-direction). Between the chips, there are gaps measuring 184~arcsec in the x-direction and 373~arcsec in the y-direction.

\begin{figure}
\begin{center}
\includegraphics[width=\columnwidth]{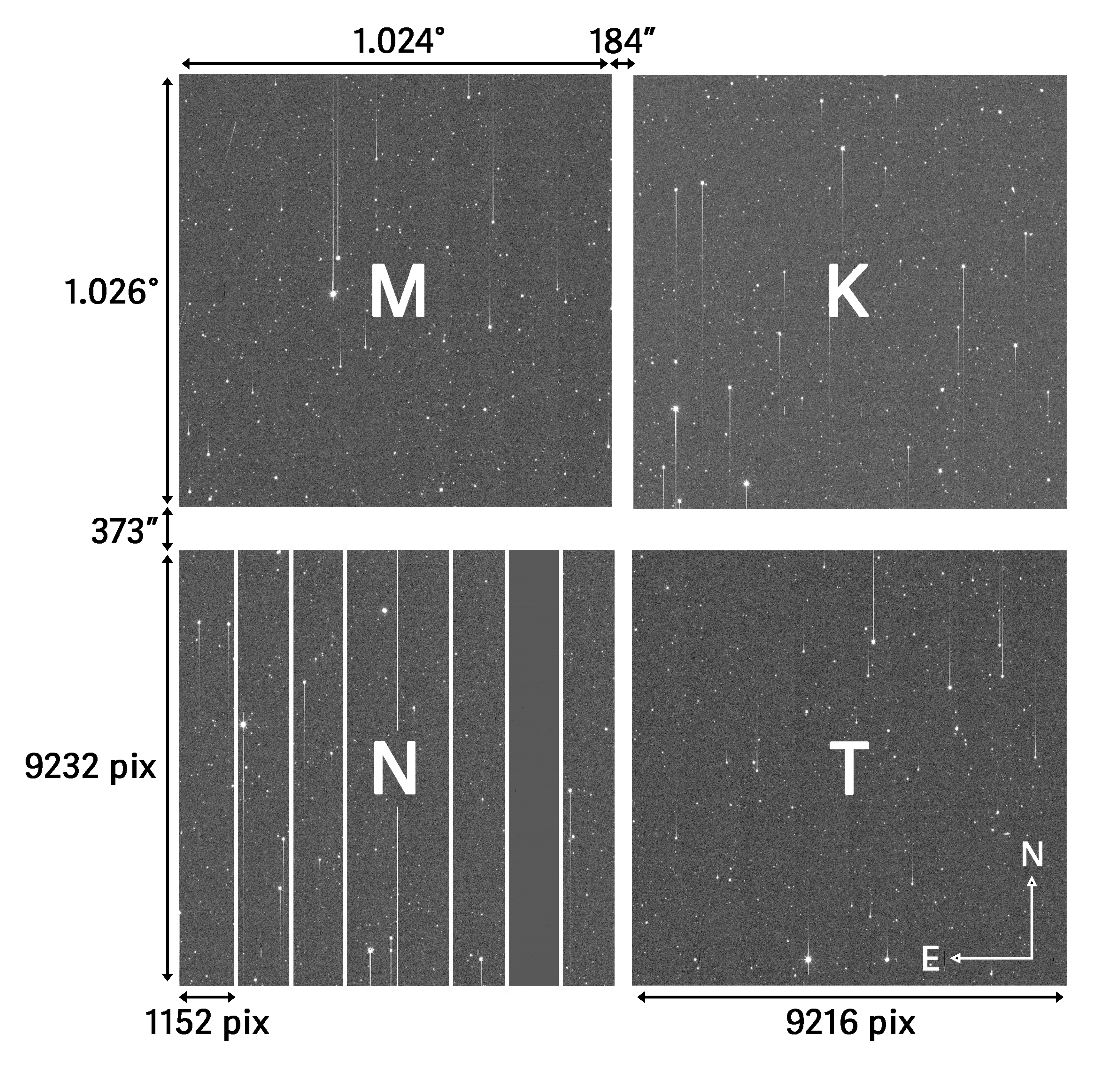}
\caption{An example of a KS4 \texttt{0001} field image observed from the CTIO observatory, illustrating the four-CCD mosaic (labeled K, M, T, and N), each measuring 9216~$\times$~9232 pixels at a scale of 0.4~arcsec~pixel$^{-1}$. Notably, the N-chip is shown subdivided into eight ``readout port images,'' each read out from a separate port. The FOV of each chip is approximately 1.024~deg~$\times$~1.026~deg, with inter-chip gaps of 184~arcsec (horizontal, east–west) and 373~arcsec (vertical, south–north). An arrow in the bottom-right corner indicates the north and east orientation.}
\label{fig:1.kmtnstructure}
\end{center}
\end{figure}

Each chip is further divided into eight ``readout port images'' (also commonly referred to as ``amplifiers'') resulting in sub-images of size 1152~$\times$~9232 pixels. Consequently, a full KMTNet image consists of 32 readout port images across the four-chip mosaic. An example illustrating the KMTNet layout is shown in Figure~\ref{fig:1.kmtnstructure}.

\subsection{KMTNet Data Acquisition and Preprocessing} \label{sec:2.2.KMTNet Data Acquisition and Preprocessing}

The KS4 data acquisition process follows a predefined tiling pattern consisting of 2,748 fields, covering declinations from $-85^\circ$ to $-30^\circ$. During the survey campaign, the KMTNet operators conducted observations according to the observing schedule that we provided. Each field was observed in four broadband filters ($BVRI$), with four 120-second exposures per filter. To fill the CCD gaps in the KMTNet mosaic camera, exposures were dithered by approximately 4~arcmin in right ascension and 7~arcmin in declination between consecutive frames. A comprehensive description of the KS4 observation strategy is provided in the KS4 overview paper (Im et al., in preparation).

The observed data are processed using the standard preprocessing pipeline developed by the Korea Astronomy and Space Science Institute (KASI), which performs overscan correction, dark subtraction, flat-fielding, and cross-talk removal \citep{2009PKAS...24...83K, 2013PKAS...28....1K}. Overscan correction is applied in place of bias subtraction due to the presence of a bias jump phenomenon in KMTNet CCDs. Cross-talk refers to electronic interference during CCD readout that produces ghost-like traces of bright sources in adjacent readout channels. In the pipeline, predefined cross-talk coefficients are assigned to each CCD and applied in batch processing to subtract the predicted cross-talk signal \citep{2016PKAS...31...35K}.

Once the standard preprocessing by the KASI pipeline is complete, the data products are ingested into the first stage of our KS4 pipeline: data quality assurance. At this stage, each image is analyzed using \texttt{Source Extractor} \citep[hereafter \texttt{SExtractor};][]{1996A.AS..117..393B} to characterize basic image properties, including source counts, background level, and seeing condition. Based on these diagnostics, images are filtered using a set of quality criteria. Those that fail to meet the standards are flagged as poor-quality and excluded from subsequent processing.

Poor-quality images are rejected based on the following three conditions. First, images exhibiting negative background values are excluded, which is a typical condition caused by readout errors. Second, images with a median full width at half maximum (FWHM) of point sources greater than 6 arcsec are rejected, as they were likely obtained under poor observing conditions and are unsuitable for use as reference frames. While a more stringent seeing cut would be ideal for reference image construction, we adopted this conservative threshold in the initial stage to prioritize maximal survey coverage and completeness. In practice, images with significantly poor seeing (e.g., FWHM $>$ 4 arcsec) were often rejected during subsequent astrometric quality control, as degraded image quality frequently prevented the pipeline from achieving the required sub-pixel positional accuracy. Third, images showing evidence of tracking errors are excluded if the average \texttt{ELONGATION} value of sources exceeds 1.9. This threshold was established using a sample of 3,700 images acquired in 2020, among which 15 were visually confirmed to exhibit tracking errors. All 15 cases were successfully rejected by this criterion, while all normal images were retained.


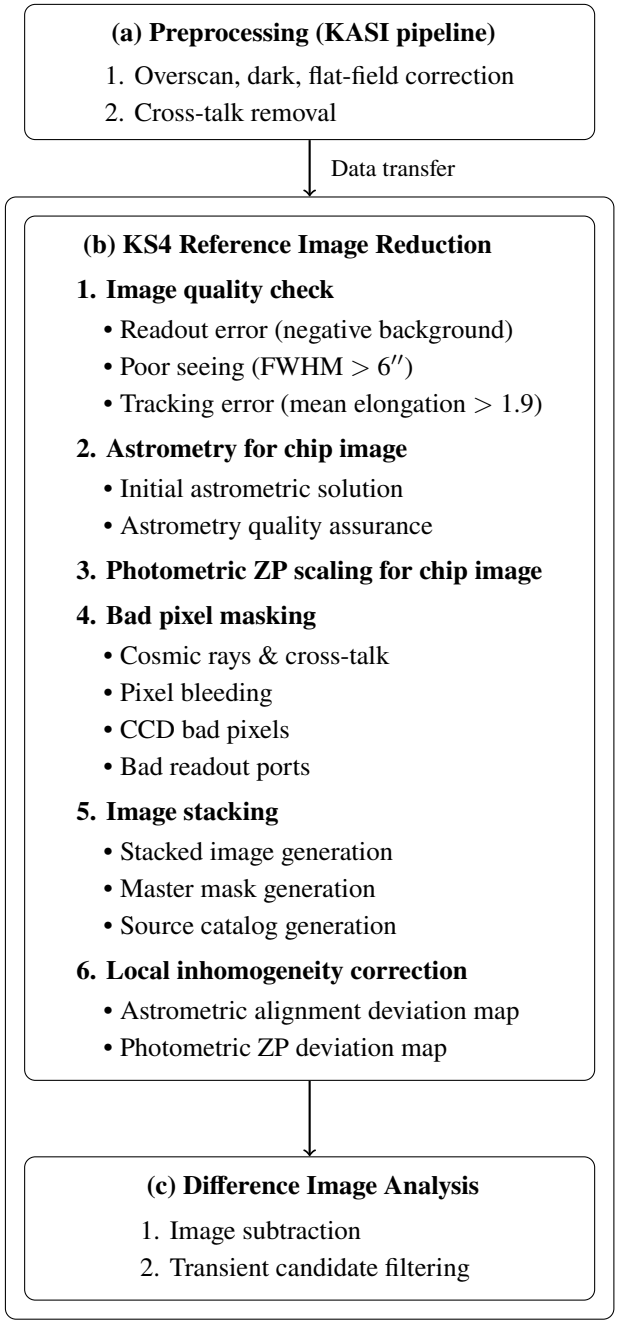
\begin{figure}[ht!]
\centering
\begin{tikzpicture}[
    node distance=10mm,
    box/.style={
        rectangle,
        draw,
        rounded corners,
        align=left,
        minimum width=7.5cm,
        inner sep=7pt
    },
    arrow/.style={->, thick}
]


\node[box] (preprocess) {
    \setstretch{1.1}
    \textbf{(a) Preprocessing (KASI pipeline)}\\[4pt]
    1. Overscan, dark, flat-field correction\\[2pt]
    2. Cross-talk removal
};

\node[box, below=of preprocess] (reference) {
    \setstretch{1.1}
    \textbf{(b) KS4 Reference Image Reduction}\\[5pt]

    \textbf{1. Image quality check}\\[3pt]
    \quad \textbullet\;Readout error (negative background)\\[2pt]
    \quad \textbullet\;Poor seeing (FWHM $>$ 6$''$)\\[2pt]
    \quad \textbullet\;Tracking error (mean elongation $>$ 1.9)\\[5pt]

    \textbf{2. Astrometry for chip image}\\[3pt]
    \quad \textbullet\;Initial astrometric solution\\[2pt]
    \quad \textbullet\;Astrometry quality assurance\\[5pt]

    \textbf{3. Photometric ZP scaling for chip image}\\[5pt]

    \textbf{4. Bad pixel masking}\\[3pt]
    \quad \textbullet\;Cosmic rays \& cross-talk\\[2pt]
    \quad \textbullet\;Pixel bleeding\\[2pt]
    \quad \textbullet\;CCD bad pixels\\[2pt]
    \quad \textbullet\;Bad readout ports\\[5pt]

    \textbf{5. Image stacking}\\[3pt]
    \quad \textbullet\;Stacked image generation\\[2pt]
    \quad \textbullet\;Master mask generation\\[2pt]
    \quad \textbullet\;Source catalog generation\\[5pt]

    \textbf{6. Local inhomogeneity correction}\\[3pt]
    \quad \textbullet\;Astrometric alignment deviation map\\[2pt]
    \quad \textbullet\;Photometric ZP deviation map
};

\node[box, below=of reference] (dia) {
    \setstretch{1.1}
    \textbf{(c) Difference Image Analysis}\\[5pt]
    1. Image subtraction\\[2pt]
    2. Transient candidate filtering
};

\node[
    draw,
    rounded corners,
    fit=(reference)(dia),
    inner sep=7pt,
    label={[font=\bfseries]above:{}}
] (groupbox) {};

\draw[arrow] (preprocess.south) -- node[midway, xshift=11mm]{\small Data transfer} (groupbox.north);
\draw[arrow] (reference) -- (dia);

\end{tikzpicture}

\caption{Flowchart of the data reduction and transient detection pipeline after image acquisition. 
\textbf{(a)} Preprocessing steps performed by the KASI pipeline \citep{2013PKAS...28....1K}. 
\textbf{(b)} KS4 DR1 reference image reduction sequence. 
\textbf{(c)} Difference image analysis (DIA) applied to newly obtained ToO images, which are reduced using the same procedures as the KS4 images and subsequently subtracted from the reference frame.}
\label{fig:pipeline}
\end{figure}

\section{Data Calibration Methods} \label{sec:3.Data Calibration Methods}

Following the initial quality assurance steps, we applied a series of calibration procedures to the KS4 images. These include astrometric and photometric calibration, as well as bad pixel masking. Final reference images were then generated through stacking, after which spatial corrections to astrometry and photometric zero-point (ZP) were applied using local deviation maps. The overall workflow of the KS4 data reduction pipeline is summarized in Figure~\ref{fig:pipeline}.

\subsection{Astrometry} \label{sec:3.1.Astrometry}

The process of obtaining precise astrometry is divided into two steps. The first step involves determining an initial astrometric solution, and the second step evaluates and ensures the accuracy of the obtained solution.

\subsubsection{Initial Astrometry} \label{sec:3.1.1.Initial Astrometry}

Astrometric calibration in the KS4 image reduction pipeline is performed using \texttt{SCAMP} \citep{Bertin2006ASPC..351..112B}, which computes a World Coordinate System (WCS) solution by matching sources detected with \texttt{SExtractor} to an external reference catalog. The key \texttt{SCAMP} parameters configured for this task are as follows:

\begin{itemize}
    \item \texttt{ASTREF\_CATALOG} is set to \texttt{UCAC-4}.
    \item \texttt{ASTREFMAG\_LIMITS} restricts the reference stars to a magnitude range of 15.0–20.0.
    \item \texttt{CROSSID\_RADIUS}, the initial matching radius, is set to 5.0 arcseconds.
    \item \texttt{POSITION\_MAXERR}, the maximum allowed positional error, is 0.5 arcminutes.
    \item \texttt{DISTORT\_DEGREES} is set to 5 to fit a fifth-degree polynomial distortion model.
    \item \texttt{PROJECTION\_TYPE} is fixed to \texttt{TPV} for compatibility with subsequent tools used in stacking and photometry.
\end{itemize}

For the initial astrometric calibration of KS4 chip images, the UCAC-4 catalog \citep{2013AJ....145...44Z} was adopted as the reference because a properly curated Gaia EDR3 catalog \citep{2021A&A...649A...2L} was not yet available during the early stage of the KS4 project. To ensure internal consistency across early reductions, all chip images in DR1 are uniformly calibrated using UCAC-4. The astrometric accuracy of the final products is later assessed against Gaia EDR3, as described in Section~\ref{sec:3.1.2.Astrometry Quality Assurance}.

We assess the quality of the resulting astrometric solution using the \texttt{ASTRRMS1} and \texttt{ASTRRMS2} keywords in the generated headers, which quantify the positional scatter in right ascension and declination relative to the reference catalog. The astrometry outcome is considered successful if both values fall below $10^{-4}$ degrees.

If this criterion is not satisfied, the \texttt{DETECT\_THRESH} parameter in \texttt{SExtractor} is increased iteratively by 10 to prioritize high signal-to-noise stars and improve matching robustness, particularly in crowded fields. Images that fail to meet the \texttt{ASTRRMS} thresholds after five iterations are excluded from further analysis, as such failures typically arise from unfavorable observing conditions, including poor seeing or cloud contamination.

\subsubsection{Astrometry Quality Assurance} \label{sec:3.1.2.Astrometry Quality Assurance}

Even after applying the astrometry process with \texttt{SCAMP}, some images exhibit inaccuracies, especially at the edges. To address this, we implemented a detailed quality assurance (QA) process using the Gaia EDR3 as the reference catalog.

Each chip image was divided into 64 segments by splitting it into 8 rows and 8 columns. Sources in each segment were matched to those in the reference catalog to evaluate the local astrometric precision. A chip was considered astrometrically reliable if at least 62 of the 64 segments satisfied both of the following criteria:

\begin{itemize}
\item At least 60\% of the reference-catalog sources in a segment are matched to detected sources within 2 arcsec.
\item The RMSE of the 3$\sigma$-clipped separation distances between matched sources is less than 0.5 arcsec.
\end{itemize}

The first criterion is typically satisfied since the image depth is sufficient to detect the majority of Gaia catalog stars in each segment, typically several tens of sources. Failure to meet this requirement usually indicates a catastrophic astrometric offset or another severe image anomaly. The second criterion assesses the precision of the astrometric solution, with the 3$\sigma$-clipping applied to reduce the influence of outliers on the RMSE calculation. We note that exceptions were allowed for up to ten segments affected by large galaxies or bright stars, where saturated regions prevented reliable source matching.

The chip images that satisfied these QA standards exhibited median RMSEs of astrometry of 0.277 arcsec, 0.290 arcsec, 0.322 arcsec, and 0.319 arcsec in $B$-, $V$-, $R$-, and $I$-band chip images, respectively. These results demonstrate that the chip images meeting the QA criteria provide reliable astrometric solutions across the KS4 fields. The results of the QA process are recorded in the image header, including the RMSE, the number of matched stars, and the encoded locations of unreliable regions. Only images meeting the QA criteria proceed to the photometric calibration and stacking process.

\subsection{Photometric Zero-Point Scaling} \label{sec:3.2.Photometric Zero-Point Scaling}

Once precise astrometry is established, photometric calibration process is conducted. A key challenge in this process is the spatial inhomogeneity of the photometric ZP across KMTNet images. This variation arises primarily from differences among CCD readout ports and imperfections in flat-field correction. To mitigate these effects, we implemented a homogenization process that normalizes the ZP to a uniform value of 30~AB mag across the field.

\begin{figure}
\begin{center}
\includegraphics[width=\columnwidth]{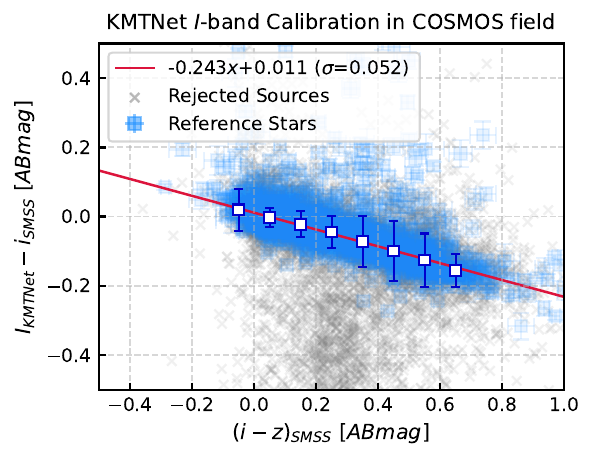}
\caption{Stellar photometry from the COSMOS field is used to calibrate the KMTNet $I$-band. The $x$-axis shows the $(i - z)$ color from the SMSS catalog, and the $y$-axis shows the difference between the KMTNet $I$-band magnitude and the SMSS $i$-band magnitude. The KMTNet $I$-band magnitudes are calibrated to the Johnson-Cousins system using known transformations in the COSMOS field. The distribution consists of the photometric reference stars (blue points) after the rejection of outliers (gray crosses), and binned statistics (white squares) representing the median and 1$\sigma$ dispersion of the reference stars for each 0.1 mag interval in $(i - z)$. The red line represents a robust linear fit to the data points.}
\label{fig:2.Icalib}
\end{center}
\end{figure}

\subsubsection{Photometric Calibration Using External Catalogs} \label{sec:3.2.1.Photometric Calibration Using External Catalogs}

\begin{figure*}
  \centering
  \includegraphics[width=\textwidth]{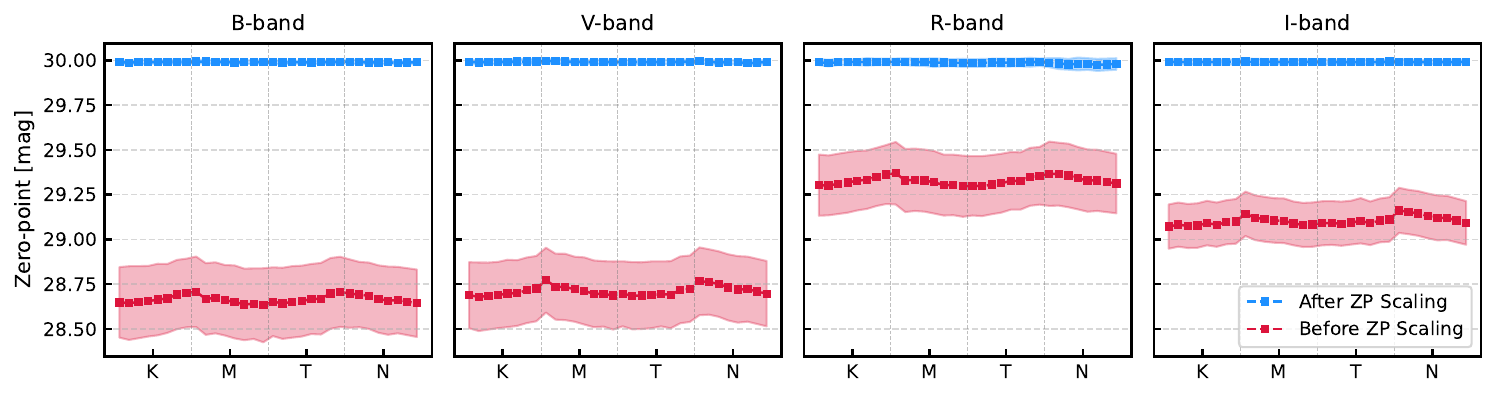}
  \caption{Photometric ZP distribution of KS4 DR1 chip images before (red) and after (blue) the ZP scaling process. Individual points represent the mean ZP for each of the 32 readout ports averaged over 979 image sets; shaded regions indicate the 1$\sigma$ standard deviation of these measurements. The pronounced scatter before the scaling process is primarily attributed to systematic ZP offsets between the different observation sites. The four panels show the distribution for the $B$, $V$, $R$, and $I$ bands from left to right.}
  \label{fig:4.chipcal}
\end{figure*}

Photometric calibration for KS4 was performed based on external reference catalogs, instead of the traditional standard star calibration. This approach requires the reference catalog to provide reliable photometry, uniform sky coverage, and filter systems suitable for accurate ZP calibration across the field. The AAVSO Photometric All-Sky Survey Data Release 9 \citep[APASS DR9;][]{2016yCat.2336....0H} catalog is particularly valuable in our cases \citep{2016yCat.2336....0H}, as they provides Johnson $B$, $V$, and Sloan $g$, $r$, $i$ magnitudes for bright stars over the entire sky. 

However, due to differences in filter transmission between APASS and KMTNet, we should apply transformation equations to convert APASS magnitudes into the KMTNet system. The calibration of the $B$-band was performed using the transformation equations proposed by \citet{2017ApJ...848...19P}, while the $R$-band calibration adopted the relation from \citet{2007AJ....133..734B}, which converts Sloan $r$- and $i$-band magnitudes into the Johnson-Cousins $R$-band. While the KMTNet $B$- and $R$-band conversions require additional color-term corrections using $(B - V)$ and $(r - i)$, no such correction was applied for the KMTNet $V$-band, as it reportedly matches the Johnson $V$-band closely. These transformations are summarized in Equations~\ref{eq:1.Bcalib}–\ref{eq:3.Rcalib}.

The $I$-band calibration was more challenging due to the absence of a suitable conversion formula from APASS photometry. To address this, we derived an empirical transformation using the SkyMapper Southern Survey Data Release 3 \citep[SMSS DR3;][]{2019PASA...36...33O}, which provides $uvgriz$ photometry across the southern sky, making it well-suited for KS4 calibration.

To derive the conversion, we utilized KMTNet data of the COSMOS field taken at SSO on 2018 December 7, where reliable Johnson-Cousins $I$-band calibration could be performed \citep{2018JKAS...51...89K}. The reference $I$-band magnitudes were estimated using Pan-STARRS $i$-band magnitudes and $(r - i)$ colors, following the method of \citet{2018BlgAJ..28....3K}. These calibrated $I$ magnitudes were then compared with SMSS $i$-band magnitudes to investigate a color-dependent transformation based on $(i - z)$.

For the calibration, we applied a set of quality cuts to select reference stars. Sources were extracted using \texttt{SExtractor}, with selections of \texttt{FLAGS = 0} and \texttt{CLASS\_STAR > 0.9}, and matched to SMSS DR3 point sources with photometric uncertainties in both $i$- and $z$-bands below 0.05 mag. A robust linear regression using \texttt{HuberRegressor} from \texttt{scikit-learn} \citep{2011JMLR...12.2825P} was then applied to model the magnitude difference $I_{\rm KMTNet} - i_{\rm SMSS}$ as a function of the $(i - z)$ color. In this process, we adopted the \texttt{MAG\_AUTO} values for the KMTNet magnitudes, while the SMSS DR3 magnitudes were based on point spread function (PSF) measurements.

The resulting fit is presented in Figure \ref{fig:2.Icalib}, showing a 1$\sigma$ scatter of 0.052 mag. This dispersion is dominated by the propagation of the mean photometric measurement errors from the reference stars ($\sigma_{ref}$) and the intrinsic scatter of the transformation equation ($\sigma_{int}$). To isolate $\sigma_{int}$, we subtracted the $\sigma_{ref}$ from the total scatter in quadrature. We found that the $\sigma_{int}$ remains largely stable across various error-cut thresholds applied to the reference stars--ranging from 0.027~mag to 0.033~mag for cuts between 0.02 and 0.05~mag--with a characteristic value of 0.031~mag. This indicates that the linear relationship is statistically robust for photometric calibration of KMTNet $I$-band images.

The final transformation equations used for chip-level calibration in KS4 are:
\begin{align}
B &= B_{\rm APASS} - 0.06 - 0.27 \times (B - V)_{\rm APASS} \label{eq:1.Bcalib} \\
V &= V_{\rm APASS} + 0.02 \label{eq:2.Vcalib} \\
R &= r_{\rm APASS} + 0.0383 - 0.3718 \times (r - i)_{\rm APASS} \label{eq:3.Rcalib} \\
I &= i_{\rm SMSS} + 0.011 - 0.243 \times (i - z)_{\rm SMSS} \label{eq:4.Icalib}
\end{align}

\noindent While the input $B_{\rm APASS}$ and $V_{\rm APASS}$ magnitudes are on the Vega system, the transformation equations (Equations~\ref{eq:1.Bcalib}--\ref{eq:4.Icalib}) are designed to convert all photometry onto a consistent AB magnitude system.

\subsubsection{Zero-point Homogenization for Chip Images} \label{sec:3.2.2.Zero-point Homogenization for Chip Images}

While the equations in Section~\ref{sec:3.2.1.Photometric Calibration Using External Catalogs} can establish a global photometric calibration for each chip, they do not account for significant spatial variations in the ZP across the detector. As illustrated by the red points in Figure~\ref{fig:4.chipcal}, these inhomogeneities must be corrected to ensure uniform photometric precision. Although the ZP variation patterns tend to be consistent within data from the same observatory, they differ noticeably between sites. Since KS4 combines observations from multiple observatories, uncorrected ZP differences can introduce significant photometric uncertainties in the final stacked images.

Therefore, homogenizing the ZP across each chip is essential for reliable photometric calibration. To assess and correct spatial ZP variations, we segmented each chip image according to its eight readout ports, which often exhibit discrete ZP offsets. For each readout port region, we measured the local ZP and evaluated its spatial variation across the image.

To estimate the ZP in each region, sources in the KS4 images were matched to their reference catalog counterparts, and the \texttt{AUTO} aperture magnitudes were compared to the reference values transformed via Equations~\ref{eq:1.Bcalib}--\ref{eq:4.Icalib}. Reliable calibration required selecting clean sources, which we achieved by applying criteria such as \texttt{FLAGS < 4} to exclude saturated or problematic detections, \texttt{MAGERR\_AUTO < 0.05} to ensure photometric precision, and \texttt{CLASS\_STAR > 0.8} to prioritize point-like sources.

To quantify and correct ZP gradients, we performed a linear regression using the \texttt{HuberRegressor} from the \texttt{scikit-learn} package \citep{2011JMLR...12.2825P}. If the linear model yielded a negative score, indicating a poor fit, we instead adopted a 2D fit using \texttt{curve\_fit} from \texttt{scipy.optimize} \citep{2020NatMe..17..261V}. After fitting, we subtracted the background level estimated by \texttt{SExtractor} and applied a multiplicative flux scaling to normalize the ZP to 30~AB mag uniformly across each chip image.

Consequently, the ZPs of the 979 KS4 DR1 images were standardized to 30~AB~mag across all 32 readout ports. Validation photometry of the scaled images confirms that systematic ZP variations between observation sites have been effectively mitigated, as illustrated in Figure~\ref{fig:4.chipcal}. While the calibration should ensure the exact 30~AB~mag ZPs, a slight scatter of $\sim$0.01~mag remains. This dispersion arises from minor discrepancies in reference star selection during validation—specifically the implementation of bad pixel masks that were not utilized during the initial homogenization.

\subsection{Bad Pixel Masking} \label{sec:3.3.Bad Pixel Masking}

\begin{figure*}
\centering
\includegraphics[width=\textwidth]{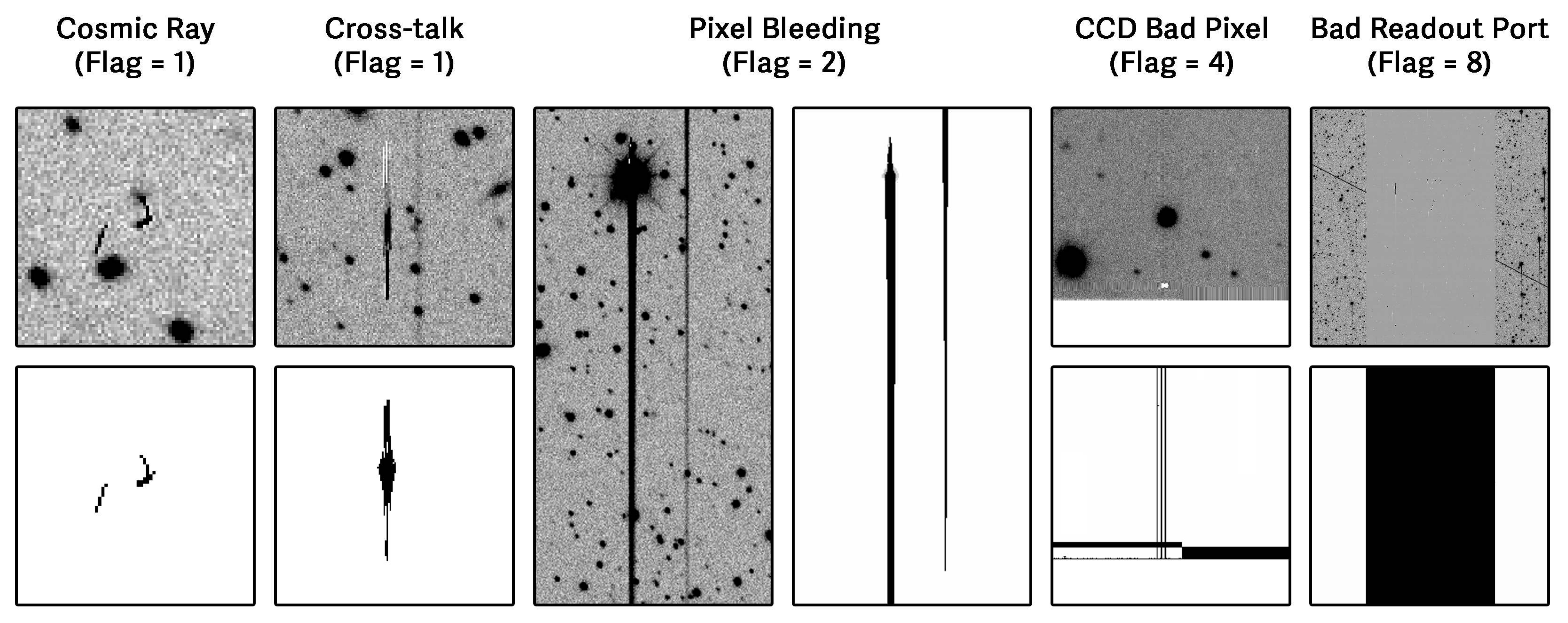}
\caption{Example cutout images illustrating the primary contaminating features flagged in the KS4 bad pixel mask. For each source type, a science image snippet (grayscale) and the corresponding mask region (binary) are shown. From left to right: cosmic rays, cross-talk artifacts, pixel bleeding from saturated stars, CCD bad pixels at the edges, and defective readout ports of the CTIO N-chip CCD. Due to their elongated nature, pixel bleeding features are displayed in an extended vertical layout.}
\label{fig:5.badmask}
\end{figure*}

KMTNet images frequently exhibit bad pixel regions caused by cosmic rays, saturated stars, and readout artifacts. These defects can systematically bias photometric measurements and lead to numerous spurious detections, particularly in difference imaging. To mitigate these effects, we constructed a bad pixel mask for each image. This mask encodes the location and type of contamination as an integer-valued array, allowing for efficient masking during image stacking, source detection, artifact flagging, and DIA.

Each contamination type is assigned a distinct bitwise value in the mask:

\begin{itemize}
\item Cosmic rays and cross-talk artifacts are labeled as 1.
\item Bleeding patterns from saturated stars are labeled as 2.
\item CCD bad pixels are labeled as 4.
\item Faulty readout ports are labeled as 8.
\end{itemize}

\noindent Example image cutouts illustrating each contamination type and the corresponding mask regions are shown in Figure~\ref{fig:5.badmask}.

\subsubsection{Cosmic Rays} \label{sec:3.3.1.Cosmic Rays}

Cosmic ray contamination is identified using \texttt{astroscrappy}, a Python implementation of the \texttt{L.A.Cosmic} algorithm \citep{2001PASP..113.1420V}. The \texttt{detect\_cosmics} function is applied to each chip image with the following parameter settings: a clipping threshold (\texttt{sigclip}) of 4.5$\sigma$ to identify pixels affected by cosmic ray, a flux fraction threshold (\texttt{sigfrac}) of 0.3, which determines the minimum relative flux that adjacent pixels must have to be considered part of the cosmic ray track, and an object contrast limit (\texttt{objlim}) of 5.0 to prevent bright astrophysical sources from being misidentified. The cleaning method is set to median-masked replacement (\texttt{cleantype = ‘medmask’}), and the detection process is iterated four times to refine the mask.

\subsubsection{Cross-talk} \label{sec:3.3.2.Cross-talk}

Cross-talk artifacts are ghost-like residual signals caused by electronic interference between readout channels during simultaneous multi-port readout \citep{2016PKAS...31...35K}. Although initial correction was applied during preprocessing of the KASI pipeline using a cross-talk coefficient defined for each observatory and CCD chip, this coefficient can vary over time. As a result, residual cross-talk artifacts often remain in the images. These residuals must be masked, as they can be mistaken for transient sources in the DIA process.

When a bright star resides in a given position within a particular readout port, it induces artifacts only in every other port—that is, stars in odd-numbered ports produce cross-talk in the other odd ports, and stars in even-numbered ports produce cross-talk in the other even ports. Within each affected port, the artifact appears at a position determined by the port’s readout orientation: the intra-port X-coordinate is mirrored about the center of the affected port. This produces a characteristic pattern of faint, horizontally aligned echoes distributed across alternating ports, with X-coordinates that are locally flipped within their respective port boundaries. (For a visual illustration of the cross-talk geometry, see Figure 1 of \citet{2016PKAS...31...35K}.) In our pipeline, we identify bright stars with peak fluxes exceeding 56,000 ADU as potential cross-talk sources and mask all predicted artifact locations accordingly.

\subsubsection{Pixel Bleeding} \label{sec:3.3.3.Pixel Bleeding}
Saturated stars in KMTNet images produce prominent vertical bleed trails caused by charge overflow. These artifacts severely corrupt the photometry of the source and nearby objects. More critically, the resulting deblending failures often lead to the exclusion of these sources from the final catalog. To mitigate these issues, it is essential to identify and mask bleeding-affected regions.

The threshold of bleeding varies across CCD chips. While the typical saturation level ranges from 50,000 to 55,000 ADU, the CTIO N-chip shows a notably lower threshold near 44,000 ADU. The bleeding direction also differs by chip: it extends downward (negative Y-direction) in the K- and N-chips, and upward (positive Y-direction) in the M- and T-chips.

Detection and masking of bleeding patterns follow a systematic procedure. First, saturated pixels are identified by thresholding above the chip-specific saturation level. For each saturated pixel, we compute the sum of pixel values within a vertical segment spanning 20--40 pixels along the expected bleeding direction, excluding background contribution. If the integrated flux exceeds 500 ADU, the segment is flagged as bleeding-affected. The extent of the bleeding is then traced by analyzing pixel values along the Y-axis until six consecutive pixels fall below 0.4$\sigma$ of the background level, marking the termination of the streak. A detailed analysis of this procedure is presented in Shin et al. (in preparation).

\subsubsection{CCD Bad Pixels} \label{sec:3.3.4.CCD Bad Pixels}

CCD bad pixels occur at fixed positions on the detectors and are specific to each observatory. These pixels are already identified and interpolated using surrounding values during preprocessing with the KASI pipeline. However, the interpolation often leads to unreliable photometry and source detection, and can generate spurious artifacts in DIA. To mitigate these effects, the same CCD bad pixel maps used in preprocessing are incorporated into the KS4 bad pixel masking procedure, ensuring consistent treatment by masking the corresponding positions and flagging affected sources.

\subsubsection{Bad Readout Port} \label{sec:3.3.5.Bad Readout Port}

A known defect in the KMTNet system involves a malfunctioning readout port. Specifically, the seventh readout port of the CTIO N-chip has been inoperative since October 2019 and remains non-functional to date (see Figure~\ref{fig:1.kmtnstructure} for an example). While this region contains no valid signal in a single exposure, it can introduce significant artifacts during the image stacking process. As dithered exposures are combined, the boundary of this empty region may overlap with valid sky areas from other frames, creating artificial flux gradients and spurious structures in the final stacked image. To prevent such artifacts and ensure photometric integrity, the entire readout port area is masked throughout the reduction pipeline.

\subsection{Image Stacking} \label{sec:3.4.Image Stacking}

In this stage of the pipeline, calibrated individual exposures are combined into deep, science-ready products. This stacking procedure employs the \texttt{SWarp} software \citep{2010ascl.soft10068B}, which performs image reprojection and pixel-wise coaddition. The process generates two distinct outputs: a stacked science image with uniform geometry, and a corresponding master bad pixel mask that combines contamination flags from the individual frames. Each stacked image has a fixed physical size of 22,000~$\times$~22,000 pixels, centered on the predefined KS4 field center. This setup ensures that the $B$, $V$, $R$, and $I$ stacked images for a given field are aligned in both pixel and WCS coordinates, enabling consistent multi-band photometry and DIA without requiring additional resampling. We incorporated a background subtraction process during the stacking, despite the individual frames being previously background-subtracted. This measure was necessary to mitigate residual background fluctuations introduced during the resampling process. The key \texttt{SWarp} parameters configured for this task are as follows:

\begin{itemize}
\item \texttt{COMBINE\_TYPE} is set to \texttt{MEDIAN} to robustly combine frames while minimizing the impact of outliers.
\item \texttt{CELESTIAL\_TYPE} is set to \texttt{EQUATORIAL} to define the coordinate system.
\item \texttt{PROJECTION\_TYPE} is set to \texttt{TAN} for tangential WCS projection.
\item \texttt{FSCALASTRO\_TYPE} is set to \texttt{VARIABLE} to account for spatial pixel scale variation across the field.
\item \texttt{RESAMPLING\_TYPE} is set to \texttt{LANCZOS3} for high-quality interpolation during resampling.
\item \texttt{SUBTRACT\_BACK} is set to \texttt{Y} (with \texttt{BACK\_SIZE} of \texttt{256} pixels) to mitigate residual background fluctuations introduced during resampling.
\end{itemize}

Bad pixel masks are combined using the same projection settings as the science images. To preserve the distinct flag values for each contamination type, we first stack the individual masks corresponding to each artifact type separately. 

The combination method varies by mask type. For cosmic rays, cross-talk, bleeding patterns, and bad readout ports, we adopt \texttt{COMBINE\_TYPE = SUM} to retain all masked regions across individual exposures. For CCD bad pixels, however, we use \texttt{COMBINE\_TYPE = MEDIAN} because these defects are fixed in detector coordinates and are generally mitigated through dithering. A median combination prevents over-masking by retaining only the consistently flagged pixels, thereby preserving the truly persistent CCD defects such as those near detector gaps.

Finally, the resulting masks are merged into a unified master mask by assigning their respective integer values, as described in Section~\ref{sec:3.3.Bad Pixel Masking}. During the final coaddition, we used \texttt{COMBINE\_TYPE = SUM}, so that pixels affected by multiple artifacts are represented by the cumulative sum of their corresponding flags.

After generating both the stacked science image and the corresponding master mask image, we apply an additional image-level cleaning process to the stacked image. This step is essential because, although bad pixel regions can be tracked using the mask, retaining contaminating sources can hinder reliable source detection. In particular, bleeding severely disrupts the deblending of nearby sources, reducing source completeness and increasing the risk of photometric bias. To mitigate these effects, we interpolated pixel values at masked locations using the surrounding pixels. This correction was performed using \texttt{SExtractor}, employing an inverted bad pixel mask as the \texttt{WEIGHT\_MAP}, where valid pixels are assigned a value of 1 and masked regions a value of 0. We adopted \texttt{MASK\_TYPE = CORRECT} and enabled \texttt{CLEAN = Y} to interpolate over masked regions and suppress spurious detections. We then generated a bad-pixel–corrected image using \texttt{CHECKIMAGE\_TYPE = -BACKGROUND}, with \texttt{BACK\_TYPE = MANUAL} and \texttt{BACK\_VALUE = 0}, producing an image identical to the original except for the interpolated regions. The resulting images were subsequently used for photometric measurements and source catalog construction.
    
\begin{figure}[h]
\begin{center}
\includegraphics[width=\columnwidth]{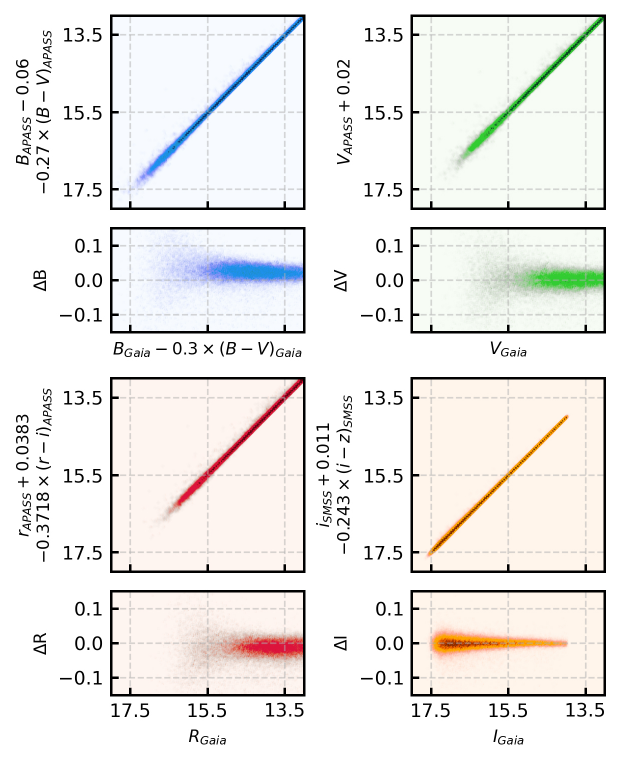}
\caption{Comparison between the previous photometric reference magnitudes (represented in Equation~\ref{eq:1.Bcalib}-\ref{eq:4.Icalib}) and the newly adopted Gaia XP magnitudes for the recalibration of KS4 stacked images. The Gaia XP magnitudes are well-calibrated to the standard Johnson-Cousins system and provide higher source completeness compared to APASS.}
\label{fig:6.Gaia_cal}
\end{center}
\end{figure}

\subsection{Source Catalog Construction} \label{sec:3.5.Source Catalog Construction}

We generated a source catalog from each stacked image using \texttt{SExtractor}. The key source detection parameters in the configuration file were:

\begin{itemize}
\item \texttt{DETECT\_MINAREA}: 9 connected pixels required for a valid detection.
\item \texttt{DETECT\_THRESH}: 1.0, specifying the detection threshold in units of background $\sigma$.
\item \texttt{ANALYSIS\_THRESH}: 1.0, specifying the analysis threshold for photometric measurement in units of background $\sigma$.
\item \texttt{DEBLEND\_NTHRESH}: 32 thresholds used to deblend overlapping sources.
\item \texttt{DEBLEND\_MINCONT}: 0.00005, defining the minimum contrast ratio for separating blended sources.
\end{itemize}

\texttt{DETECT\_MINAREA} and \texttt{DETECT\_THRESH} were chosen to suppress small-scale artifacts while maintaining high completeness for faint sources. For source deblending, we adopted a low value for \texttt{DEBLEND\_MINCONT} and a high value for \texttt{DEBLEND\_NTHRESH}, enabling reliable separation of extended or overlapping sources through fine sampling of intensity levels. Another parameter that affected source detection was \texttt{MEMORY\_PIXSTACK}, which was set to $10^7$ pixels to prevent pixel stack overflow.

To ensure accurate classification and flagging in the source catalog, several configuration parameters must be carefully specified. The point-source FWHM was estimated from representative regions within each image and assigned to the \texttt{SEEING\_FWHM} parameter for consistent application across the full field. The \texttt{SATUR\_LEVEL} was defined as the product of the minimum ZP scaling factor used in chip-level flux calibration (see Section~\ref{sec:3.2.2.Zero-point Homogenization for Chip Images}) and 60,000 ADU. For quality control, the stacked bad pixel mask was supplied as the \texttt{FLAG\_IMAGE}, enabling \texttt{IMAFLAGS\_ISO} to flag sources affected by contaminating artifacts.

With this configuration, we constructed a source catalog containing positional, photometric, morphological, and flagging information for detected objects. Further details will be provided in the forthcoming KS4 DR1 catalog paper \citep{2026JKAS...Chang}.

\subsection{Correction for Local Inhomogeneity} \label{sec:3.6.Correction for Local Inhomogeneity}

\begin{figure*}[t]
\centering
\includegraphics[width=\textwidth]{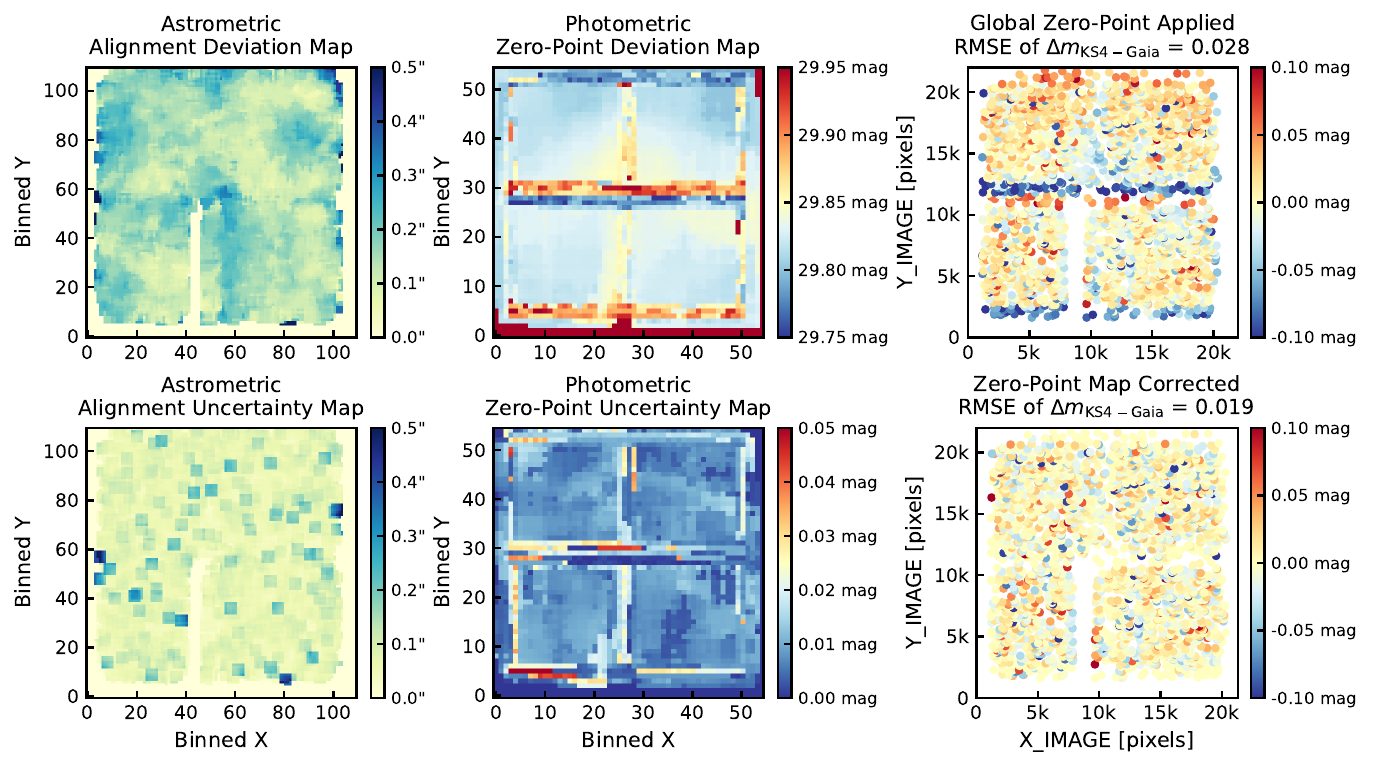}
\caption{Two-dimensional astrometric and photometric calibration results for KS4 field \texttt{0001} in the $I$-band.
\textbf{Left column:} Astrometric alignment deviation map (top) showing the average angular separation between matched KS4 and Gaia source coordinates, and the corresponding alignment uncertainty map (bottom) representing the standard deviation within each 200~$\times$200 pixel bin (Section~\ref{sec:3.6.1.Local Astrometric Alignment Deviation Map}).
\textbf{Middle column:} Photometric ZP deviation map (top) and the associated ZP uncertainty map (bottom), constructed using 400~$\times$400 pixel bins (Section~\ref{sec:3.6.2.Local Photometric Zero-Point Deviation Map}).
\textbf{Right column:} Spatial distribution of the magnitude difference between KS4 and Gaia sources in the 14--19mag range. The top panel applies a single global ZP across the field, while the bottom panel applies a spatially varying ZP correction based on the map shown in the middle column. This correction reduces the RMSE of $\Delta m_{\rm KS4 - Gaia}$ from 0.028mag to 0.019~mag.
\label{fig:7.calmap}
}
\end{figure*}

Even though the individual chip image zero-points were homogenized during the early calibration stage (Section~\ref{sec:3.2.2.Zero-point Homogenization for Chip Images}), the stacked image still exhibited significant spatial variations in ZP. These inconsistencies became particularly pronounced when combining exposures taken under different seeing conditions. Additionally, sources located near CCD gap regions showed substantial ZP fluctuations—especially when measured with smaller apertures—suggesting that the discrepancies stemmed from variations in the PSF across the field. These findings highlight the need for an additional spatial calibration step to correct such inhomogeneities in the stacked image.

To investigate this issue, we examined how astrometric alignment and photometric ZP varied across different regions of the stacked images. However, as we needed to assess the statistics in spatially narrow areas such as CCD gaps, the previously used APASS catalog was not ideal due to its limited depth and sparse coverage. Instead, we adopted the Gaia XP catalog as the new reference, which offers a deeper limiting magnitude and a higher spatial density of reference stars, making it better suited for tracing fine-scale variations across the field \citep{2023A&A...674A...1G}.

Gaia XP provides synthetic photometry calibrated to the Johnson-Cousins $BVRI$ system. This photometry shows good agreement (RMSE $\sim$0.02 mag) with the magnitude transformations derived from APASS and SMSS, which were used for individual chip-level calibrations, as shown in Figure~\ref{fig:6.Gaia_cal}. The $B$-band color term coefficient was slightly adjusted from 0.27 to 0.30 to minimize residual scatter against KS4 magnitudes. Throughout the evaluation of the local deviation maps, the 5 arcsec fixed aperture magnitudes were used as the KS4 magnitude.

\subsubsection{Local Astrometric Alignment Deviation Map} \label{sec:3.6.1.Local Astrometric Alignment Deviation Map}

To evaluate local photometric precision, we verified the astrometric accuracy in the corresponding regions. As described in Section~\ref{sec:3.1.2.Astrometry Quality Assurance}, our chip-level astrometric QA process ensures subpixel-level precision in most stacked images. However, because the quality criteria allow minor deviations in up to two of the 64 image segments, marginal astrometric inaccuracies may occur in edge regions, although these are rare exceptions. To identify and exclude such areas, we constructed a two-dimensional map that quantifies local deviations between KS4 source positions and Gaia DR3 \citep{2023A&A...674A...1G}, effectively characterizing the astrometric alignment quality across the field.

This deviation map was generated using an overlapping binning scheme designed to balance spatial resolution with statistical robustness. Specifically, each KS4 stacked image was divided into a grid of 200~$\times$~200 pixel bins. For each bin, astrometric statistics were computed from sources located within a centered 1000~$\times$~1000 pixel region. This approach enables the detection of local astrometric deviations while ensuring a sufficient number of reference stars are included in each measurement.

We selected point sources satisfying \texttt{FLAGS = 0}, \texttt{IMAFLAGS\_ISO = 0}, and \texttt{MAGERR\_APER < 0.05}, and cross-matched them to Gaia DR3 stars with $G$-band magnitude in the 14--20~mag range using a 2~arcsec radius. For each bin, we then computed the mean separation and its standard deviation for the matched pairs, producing a local astrometric alignment deviation map and the corresponding uncertainty map.

As an illustration, the left column of Figure~\ref{fig:7.calmap} presents the spatial distribution of the mean and standard deviation of separation distances between KS4 and Gaia sources. Most regions exhibit astrometric alignment better than 0.4~arcsec. However, larger deviations tend to occur near the corners of chip images—corresponding to either the outer edges or central gap regions in the stacked image—where astrometric accuracy can degrade due to marginal chip-level calibration. We note that some of the elevated deviations may also result from a locally insufficient number of matched reference stars. This interpretation is supported by the alignment uncertainty map in the lower-left panel, where regions with greater deviation often correspond to areas of increased uncertainty.

\subsubsection{Local Photometric Zero-Point Deviation Map} \label{sec:3.6.2.Local Photometric Zero-Point Deviation Map}

Evaluating local variations in the photometric ZP is more challenging than assessing astrometric accuracy, due to potential biases from faint stars with large photometric uncertainties. To mitigate this, we constructed a photometric ZP deviation map using a similar overlapping binning scheme at coarser resolution. Specifically, each KS4 image was divided into a grid of 400~$\times$~400 pixel bins, and for each bin, ZP statistics were measured from sources within a centered 4000~$\times$~4000 pixel region. This ensured a sufficient number of bright reference stars for robust estimation.

Reference stars were selected from Gaia XP in the 14--17~mag range for each band, chosen to be sufficiently bright to provide robust zero-point estimates. To ensure high-quality matches, we applied the same source-quality cuts as in the astrometric analysis (\texttt{FLAGS = 0}, \texttt{IMAFLAGS\_ISO = 0}, \texttt{MAGERR\_APER < 0.05}). A nominal matching radius of 0.5 arcsec was adopted; in regions where the local astrometric deviation exceeded this value (from the astrometric deviation map), the local deviation was used instead. Within each bin, the ZP was computed as the median KS4–Gaia XP magnitude difference after 2$\sigma$ outlier clipping, and the standard deviation of the clipped residuals was recorded as the local ZP uncertainty.

While this approach captures large-scale ZP variations, it is less sensitive to sharp discontinuities near CCD gaps. To address this, we computed an additional, gap-focused ZP map using non-overlapping 400~$\times$~400 pixel bins. When a bin lacked sufficient reference stars, its ZP was interpolated from the nearest valid bin. The resulting gap-region map was then merged with the global map to produce the final two-dimensional ZP deviation map. The complete set of calibration maps is publicly available on Zenodo\footnote{\url{https://doi.org/10.5281/zenodo.17336382}}.

Using this final map, we applied position-dependent flux scaling to each KS4 stacked image to homogenize the photometric ZP across the field, enforcing a 5 arcsec aperture ZP of 30~AB mag in all bins. An illustration of the ZP map and the resulting improvement in photometric residuals is shown in the last two columns of Figure~\ref{fig:7.calmap}. These corrected images constitute the final KS4 DR1 reference images released to the community.

\section{Result} \label{sec:4.Result}

\subsection{Stacked Image Statistics} \label{sec:4.1.Stacked Image Statistics}

\begin{figure*}
  \centering
  \includegraphics{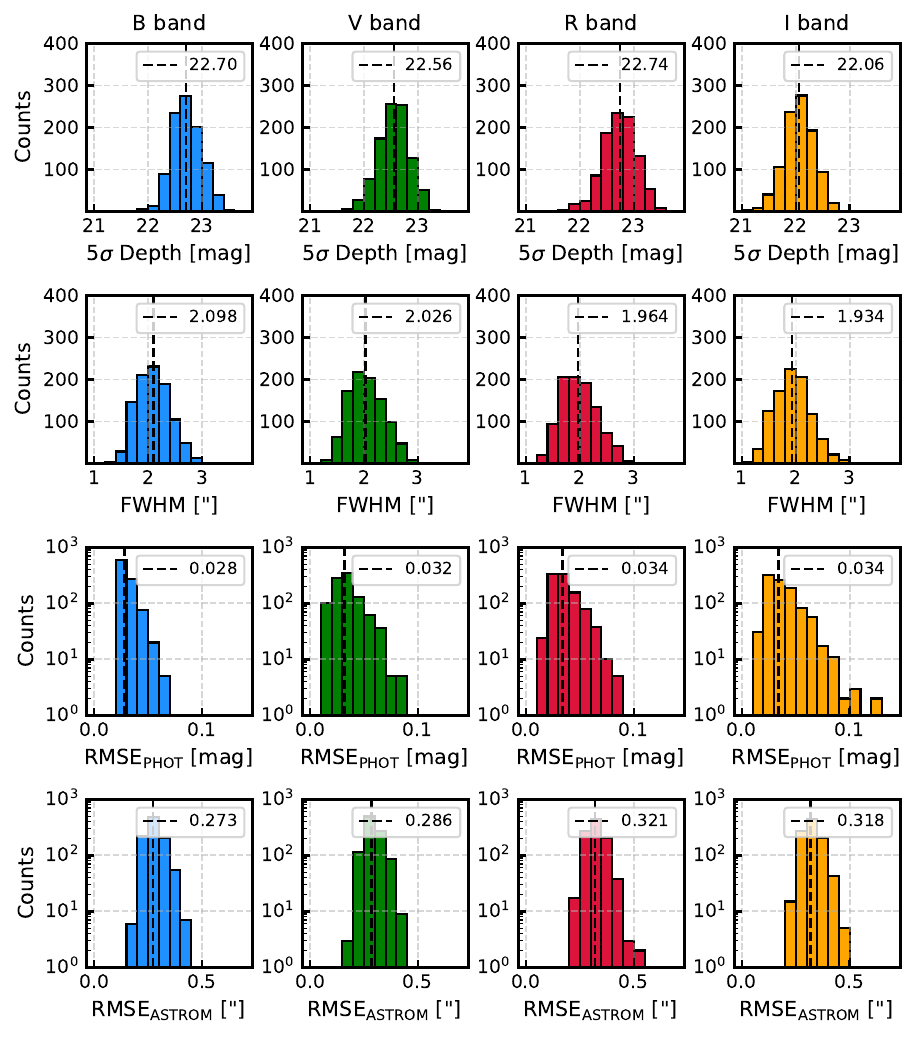}
  \caption{Statistical overview of the KS4 DR1 dataset (979 fields in each of the $B$-, $V$-, $R$-, and $I$-bands). From top to bottom, the rows show the distributions of the 5$\sigma$ limiting magnitudes, the median seeing of point sources, and the photometric and astrometric RMSE (relative to Gaia, for sources in the 14--19 mag range), respectively. Vertical dashed lines indicate the median values for each distribution, labeled accordingly. Note the logarithmic scale used for the RMSE panels. A small number of fields in the $I$-band exhibit photometric RMSE values larger than 0.1 mag due to severe source deblending challenges in areas of high stellar density.}
  \label{fig:8.overallstats}
\end{figure*}

This section summarizes the statistical properties of the KS4 DR1 stacked images after all calibration steps. The KS4 DR1 comprises 3,916 images covering 979 fields in the $B$-, $V$-, $R$-, and $I$-bands.

\subsubsection{Astrometry Statistics} \label{sec:4.1.1.Astrometry Statistics}

To evaluate astrometric precision, we used the Gaia DR3–matched statistics described in Section~\ref{sec:3.6.1.Local Astrometric Alignment Deviation Map}. For each stacked image, the field-wide RMSE of the KS4–Gaia positional separations was adopted as the representative astrometric accuracy. Across the 979 fields, the median RMSEs for the $B$-, $V$-, $R$-, and $I$-bands are 0.273 arcsec, 0.286 arcsec, 0.321 arcsec, and 0.318 arcsec, respectively. These stacked image astrometry statistics are comparable to those of the chip images, indicating that the stacking process preserves astrometric reliability. The full distributions are shown in the fourth row of Figure~\ref{fig:8.overallstats}.

In general, the $R$ and $I$ bands exhibit slightly higher astrometric RMSEs than the $B$ and $V$ bands. Images with RMSE $>$ 0.4~arcsec occur primarily in high–stellar-density fields, where crowding and deblending complicate centroiding and degrade the astrometric solution; these effects are more pronounced in the longer-wavelength bands.

Overall, the astrometric precision of the KS4 stacked images is subpixel relative to the KMTNet pixel scale and is reliable for the vast majority of the dataset. Localized regions with reduced accuracy can be identified using the astrometric deviation map introduced in Section~\ref{sec:3.6.1.Local Astrometric Alignment Deviation Map}.

\subsubsection{Photometry Statistics} \label{sec:4.1.2.Photometry Statistics}

To evaluate the photometric quality of the KS4 stacked images, we measured for each image the 5$\sigma$ limiting magnitude, the average point-source FWHM, and the photometric RMSE with respect to the external reference catalog. The distributions of statistics are presented in Figure~\ref{fig:8.overallstats}.

We estimated the 5$\sigma$ depth from the \texttt{BACKGROUND\_RMS} check image produced by \texttt{SExtractor}. For each image, the median pixel value of the \texttt{BACKGROUND\_RMS} map was adopted as the sky fluctuation. Using the standard relation between background RMS and signal-to-noise ratio, together with a seeing-sized aperture and a photometric ZP of 30~AB mag, we computed the 5$\sigma$ limiting magnitude. The median depths are 22.72, 22.56, 22.75, and 22.07~AB mag in the $B$-, $V$-, $R$-, and $I$-bands, respectively, when measured using a seeing-sized aperture.

The FWHM was measured from point sources selected with \texttt{FLAGS = 0}, \texttt{IMAFLAGS\_ISO = 0}, \texttt{MAGERR\_APER < 0.05}, and \texttt{CLASS\_STAR > 0.8}. For each image, we adopted the average of \texttt{FWHM\_WORLD} as the representative value. The median FWHM across the survey is approximately 2 arcsec, with only a small fraction of images exceeding 3 arcsec. Seeing conditions were comparable across all four bands ($BVRI$).

Additionally, we cross-matched KS4 sources to Gaia XP and measured the photometric scatter for stars in the 14--19mag range, using 5 arcsec fixed-aperture magnitudes. The resulting RMSEs are 0.026, 0.023, 0.025, and 0.026~mag for the $B$-, $V$-, $R$-, and $I$-bands, respectively. A small number of $R$- and $I$-band images show larger RMSEs, primarily in high–stellar-density fields where crowding and deblending complicate photometry.

The photometric statistics outlined above are slightly shallower than the initially planned KS4 depths of 23.5, 23.2, 23.2, and 22.5~AB mag in the $B$-, $V$-, $R$-, and $I$-bands, respectively. This shortfall can be attributed primarily to less favorable seeing conditions compared to the expected $\sim$1.6~arcsec.

Still, as a wide-area legacy imaging dataset, KS4 catalog offers significant utility for a broad range of science such as quasar discovery \citep{2024ApJS..275...46K} or galaxy cluster science (Park et al. in preparation). For detailed photometric refinements of the KS4 source catalog, refer to the companion catalog paper \citep{2026JKAS...Chang}.

\begin{table*}[t]
\centering
\caption{Summary of KMTNet ToO Observations and KS4 Coverage for Gravitational-Wave Events}
\setlength{\tabcolsep}{6pt}
\label{tab:1.gw_summary}
\begin{tabular}{lccccccc}
\toprule
\multirow{2}{*}{GW Event} & \multicolumn{3}{c}{Analyzed Area [deg$^2$]} & \multicolumn{2}{c}{5$\sigma$ Depth [AB mag]} & \multirow{2}{*}{\makecell{Transient\\Identifications}} & \multirow{2}{*}{\makecell{Reference\\(see Notes)}} \\
\cmidrule(lr){2-4} \cmidrule(lr){5-6}
& GW Area & ToO Coverage & KS4 Overlap & $R$ & $I$ & \\
\midrule
S230518h  & 460 & 224 & 160 & $21.50\pm0.70$ & -- & 5 & (1), (2) \\
S240422ed & 259 & 240 & 76  & $21.49\pm0.74$ & $21.74\pm0.19$ & 10 & (3), (4), (5) \\
S240915b  & 18  & 20  & 20  & $21.95\pm0.52$ & $21.24\pm0.71$ & 3 & -- \\
S250206dm & 547 & 80  & 48  & $21.66\pm0.25$ & $21.48\pm0.28$ & 8 & (6) \\
S250830bp & 3.73 & 20  & 20  & $21.99\pm0.24$ & $21.64\pm0.09$ & 4 & -- \\
\bottomrule
\end{tabular}

\vspace{0.5em}
\begin{minipage}{0.95\textwidth}
\small
\textbf{Notes.}
{``GW Area'' refers to the 90\% credible region reported by the LVK, while ``ToO Coverage'' is the total area observed by KMTNet (assuming 4~deg$^2$ per field). ``KS4 Overlap'' is the portion of the ToO coverage where KS4 reference images were available for DIA, and ``Transient Identifications'' counts the number of visually confirmed transients detected in at least two epochs. All 5$\sigma$ depths are average limiting magnitudes for each campaign. Detailed references for each follow-up are provided in the ``Reference'' column: 
(1) \citealt{2023GCN.33833....1P}, 
(2) \citealt{2025ApJ...981...38P}, 
(3) \citealt{2024GCN.36343....1J}, 
(4) \citealt{2024GCN.36371....1P}, 
(5) \citealt{2025arXiv250315422P}, 
(6) \citealt{2025GCN.39241....1P}.}
\end{minipage}
\end{table*}

\subsection{Scientific Validation: The Real-Time Transient Discovery Pipeline} \label{sec:4.2.Scientific Validation: The Real-Time Transient Discovery Pipeline}

The photometric depth and image quality achieved in KS4 provide data sufficiently deep for the original goal of identifying optical counterparts to GW events. For instance, the expected peak brightness of a kilonova in the $R$-band reaches approximately 21--22~AB mag at a distance of 100~Mpc, depending on the ejecta mass and composition \citep{2017Natur.551...80K}. Given the current sensitivity of GW detectors, the KS4 median depth of $R$ = 22.75~AB mag is capable of capturing a substantial fraction of bright kilonovae. Therefore, the KS4 reference images can greatly support the early recognition of kilonova candidates in KMTNet observations obtained at suitable epochs.

To fully utilize the potential of KS4 as a reference image database, we developed a system capable of performing DIA between newly acquired science images and pre-constructed reference frames during ToO observations. This automated pipeline continuously monitors incoming KMTNet data, processes the images, and identifies transient candidates in near real time.

Table~\ref{tab:1.gw_summary} summarizes five GW events from the LVK Collaboration’s fourth observing run (O4) that were followed up by KMTNet under its allocated ToO program. Their key properties are summarized below:
\begin{itemize}
\item \textbf{S230518h:} A high-probability neutron star--black hole (NSBH) merger candidate with a 90\% localization area of 460~deg$^2$ and an estimated luminosity distance of $204 \pm 57$~Mpc \citep{2023GCN.33816....1L}.
\item \textbf{S240422ed:} Initially identified as an NSBH candidate but later reclassified as a low-significance terrestrial event. The 90\% localization area spans 259~deg$^2$, with an estimated distance of $188 \pm 43$~Mpc \citep{2024GCN.36240....1L}.
\item \textbf{S240915b:} A distant binary black holes (BBH) or NSBH merger candidate ($872 \pm 149$~Mpc) with a compact 90\% localization area of 18~deg$^2$ \citep{2024GCN.37513....1L}.
\item \textbf{S250206dm:} An NSBH or binary neutron stars (BNS) merger candidate with an estimated luminosity distance of $373 \pm 104$~Mpc and a 90\% localization area of 547~deg$^2$ \citep{2025GCN.39231....1L}.
\item \textbf{S250830bp:} A distant BBH or NSBH candidate ($427 \pm 69$~Mpc) with a compact 90\% localization area of 3.73~deg$^2$ \citep{2025GCN.41607....1L}.
\end{itemize}

In each case, KMTNet performed prompt tiling of the GW localization area, and DIA was conducted using KS4 reference frames where available. These campaigns demonstrate the operational readiness of the KS4 DIA pipeline in supporting real-time follow-up of diverse GW events.

\subsubsection{Image Subtraction} \label{4.2.1.Image Subtraction}

Image subtraction between the science and reference images was performed using the \texttt{HOTPANTS} software \citep{2015ascl.soft04004B}. This program enables accurate subtraction by matching the PSF and flux levels of the two input images, allowing transient signals to be reliably identified.

Because the KMTNet images cover a wide FOV, the PSF can vary significantly across each chip. Applying a single convolution kernel to the full image can therefore lead to inadequate PSF matching. To mitigate this, each image is divided into a 4~$\times$~4 grid, and \texttt{HOTPANTS} independently derives a convolution kernel and performs subtraction in each subregion. This approach allows the kernel solution to adapt to local PSF variations across the image.

To derive an accurate convolution kernel, \texttt{HOTPANTS} requires the positions of point sources that appear in both the science and reference images. These stars serve as the reference point sources (“stamps”) for PSF-matching kernel estimation. For each subregion, we construct a stamp catalog by cross-matching the KS4 reference catalog with sources detected in the science image and selecting point sources that satisfy the following quality criteria in both catalogs: \texttt{FLAGS = 0}, \texttt{IMAFLAGS\_ISO = 0}, \texttt{CLASS\_STAR > 0.8}, and \texttt{14 < MAG\_AUTO < 20}. The coordinates of these stamps are supplied to \texttt{HOTPANTS}, which uses their image cutouts to solve for the optimal convolution kernel in each subregion.

The direction of convolution is determined by comparing the average FWHM values of the science and reference images. The image with better seeing (smaller FWHM) is convolved to match the broader PSF of the other image, following the \citet{1998ApJ...503..325A} algorithm implemented in \texttt{HOTPANTS}.

Regions flagged as bad pixels (Section~\ref{sec:3.3.Bad Pixel Masking}) and outer margins introduced during image standardization are excluded from both kernel estimation and subtraction. This prevents invalid pixels from contaminating the kernel solution or generating artifacts in the final difference image. The remaining valid regions undergo convolution and subtraction, yielding the final difference image used for transient detection.

\begin{figure}[t]
  \centering
  \includegraphics[width=\columnwidth]{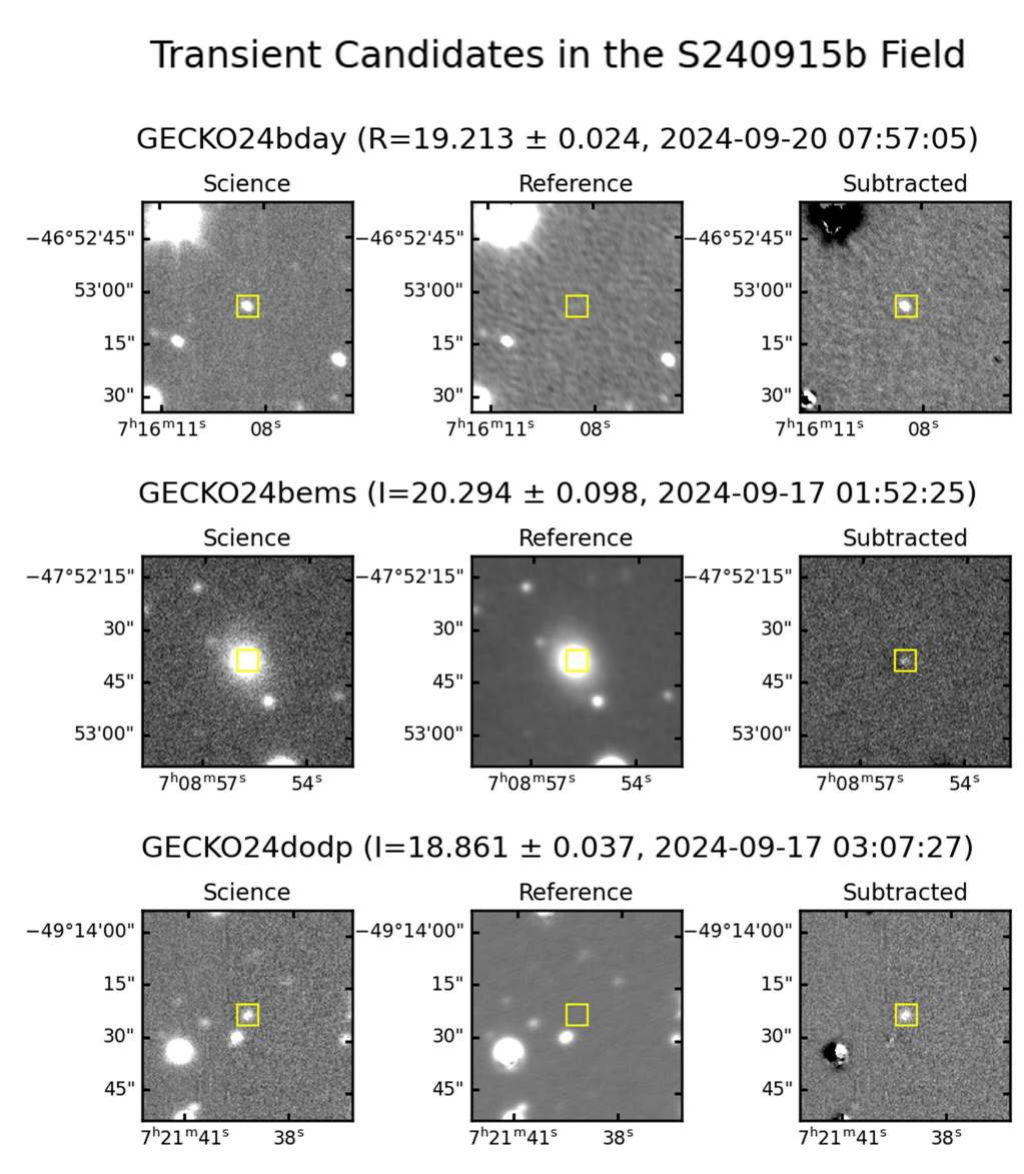}
  \caption{Transient candidates identified during the KMTNet follow-up campaign for the GW event S240915b. Each row shows a set of image cutouts corresponding to a single candidate: (1) the science image obtained during the follow-up, (2) the corresponding KS4 reference image, and (3) the resulting subtracted image. All candidates were detected at least twice and are annotated with the time of first detection and their photometric measurements.} \label{fig:9.s240915b}
\end{figure}

\begin{figure}
  \centering
  \includegraphics[width=\columnwidth]{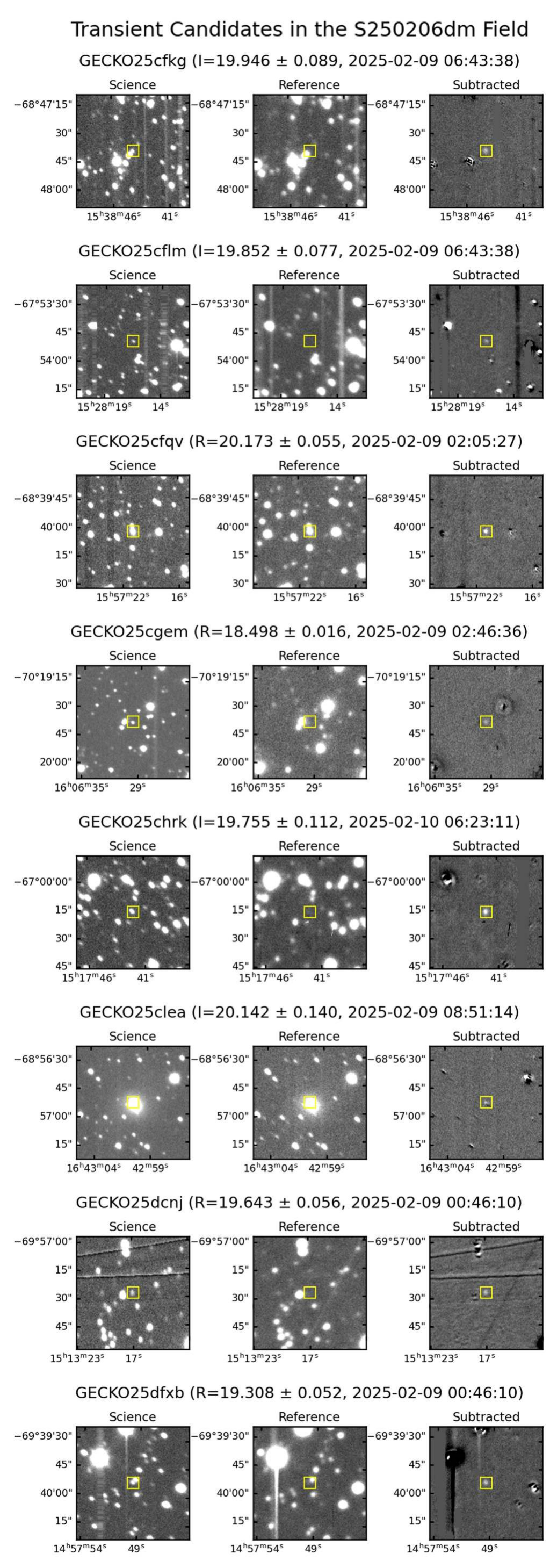}
  \caption{Same format as Figure~\ref{fig:9.s240915b}, but showing transient candidates identified during the KMTNet follow-up of GW event S250206dm.} \label{fig:10.s250206dm}
\end{figure}

\begin{figure}[t]
  \centering
  \includegraphics[width=\columnwidth]{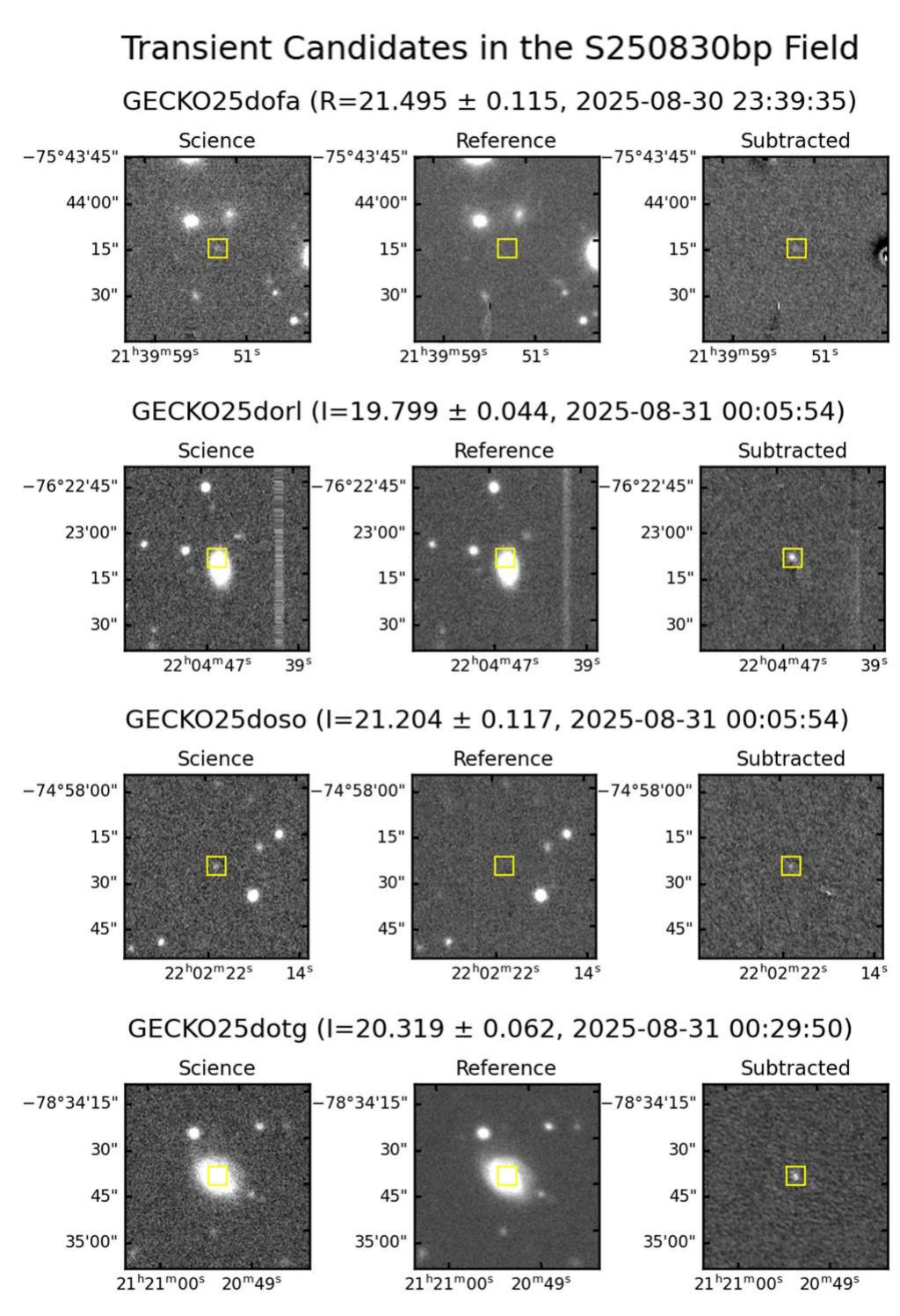}
  \caption{Same format as Figure~\ref{fig:9.s240915b}, but showing transient candidates identified during the KMTNet follow-up of GW event S250830bp.} \label{fig:11.s250830bp}
\end{figure}

\subsubsection{Filtering of Transient Candidates} \label{4.2.2.Filtering of Transient Candidates}

Subtracted images contain both astrophysical transients and spurious residuals, the latter primarily introduced by imperfections in the PSF matching process. These residuals often exhibit morphological and PSF characteristics inconsistent with genuine point sources. To discriminate between astrophysical signals and image artifacts, we apply a set of algorithmic filters based on criteria detailed below.

\begin{enumerate}
\item Positional coincidence within 5 arcsec of a known asteroid from \texttt{SkyBoT} \citep{2006ASPC..351..367B}.
\item Spatial match to a detection in the inverted subtracted image within one FWHM radius.
\item Flagged as saturated or located within known bad pixel regions (\texttt{FLAGS} > 4 or \texttt{IMAFLAGS\_ISO} > 0).
\item Elongation greater than four times the median elongation of point sources in the science image.
\item FWHM less than 0.7 or greater than 2.4 times the median FWHM of point sources in the science image.
\item Local background deviating by more than $2\sigma$ from the global median background level.
\item Location in regions where the \texttt{HOTPANTS} convolution $\chi^2$ exceeds 1000 or is a $3\sigma$ outlier.
\item Coincidence with the predicted cross-talk artifact positions.
\item Absolute value of \texttt{SPREAD\_MODEL} exceeding 1.0, indicative of non-point-like morphology (applied optionally; see main text).
\end{enumerate}

Any source in the subtracted image that satisfies one or more of these criteria is flagged as an artifact and excluded from further analysis. This filtering procedure follows the general approach of \citet{2025ApJ...981...38P} but has been streamlined and optimized to align with the data format and processing flow of the present pipeline. Among the criteria, \texttt{Criteria 2} and \texttt{7} are particularly effective at identifying artifacts associated with regions of poor PSF convolution. Meanwhile, \texttt{Criteria 4--5} further refine the sample by filtering out sources whose morphological properties—such as elongation and FWHM—deviate from the expected PSF of stellar sources in the science image.

To evaluate the efficiency of this filtering process, we generated simulated datasets by injecting synthetic point sources into blank-sky regions. Applying \texttt{Criteria 1--8} retained approximately 97\% of synthetic sources used for validation while rejecting about 81\% of artifacts. The remaining 3\% of synthetic point sources were primarily excluded because they overlapped known bad pixel regions and were subsequently removed by \texttt{Criterion 3}.

After this initial filtering, each exposure typically yields on the order of $10^3$ transient candidates. \texttt{Criterion 9} can further eliminate up to 90\% of the remaining spurious sources using \texttt{PSFEx} \citep{2011ASPC..442..435B}. However, due to the computational expense of PSF fitting for all candidates, we instead adopt a convolutional neural network (CNN)-based real–bogus (RB) classifier as a more efficient and scalable alternative when available \citep{lee2025investigatingeffectspointsource}.

Generally, the RB classifier retains only about 0.5--1\% of the initial candidates with a predicted probability greater than 50\% of being a real transient, effectively leaving a manageable number of sources for visual inspection. Each transient candidate is assigned a unique identifier following the naming conventions of the Gravitational-wave Electromagnetic Counterpart Korean Observatories (GECKO) project \citep{2020grbg.conf...25I}. The final number of confirmed transient targets is summarized in Table~\ref{tab:1.gw_summary} and Figures~\ref{fig:9.s240915b}--\ref{fig:11.s250830bp}. These candidates include both events previously reported by other surveys and newly identified sources unique to our follow-up observations. For instance, GECKO25dcnj, GECKO25cfqv, and GECKO25cflm, detected during the S250206dm campaign, correspond to AT2024aerp \citep{2024TNSTR5027....1H}, AT2025bmx \citep{2025TNSTR.594....1H}, and AT2024abkv \citep{2024TNSTR4528....1H} listed in the Transient Name Server (TNS)\footnote{\url{https://www.wis-tns.org/}}, respectively. Likewise, GECKO25dorl, discovered during the S250830bp campaign, was also confirmed as AT2025vom \citep{2025TNSTR3390....1T}.

\section{Discussion} \label{sec:5.Discussion}

Through the development of the image reduction pipeline, we have achieved high precision in both astrometric and photometric calibrations. Furthermore, our transient follow-up observations have demonstrated the value of the KS4 dataset as a legacy-quality reference image archive. Nonetheless, several aspects of the pipeline can still be improved to further enhance the precision, consistency, and overall quality of the final data products. In the following section, we outline potential future upgrades to the astrometric calibration, photometric calibration, and image stacking procedures, as well as the current limitations of this data release.

\subsection{Astrometric Calibration Improvements} \label{sec:5.1.Astrometric Calibration Improvements}

The current astrometric calibration procedure consists of three main steps: (1) deriving an initial astrometric solution using \texttt{SCAMP}, (2) performing a QA using the Gaia catalog, and (3) constructing a local deviation map to evaluate astrometric precision across the field. The first two steps are applied at the individual chip image level, while the final step is performed on the stacked images and associated source catalogs.

One notable limitation of the initial astrometric solution is the use of the UCAC-4 catalog as the reference. A more effective approach would be to adopt the Gaia DR3 catalog from the outset, as is already done in the QA and further steps. Using pre-defined Gaia catalog would not only streamline the processing pipeline but also improve the accuracy of source selection by incorporating proper motion, parallax and other types of error from Gaia. 

Another proposed improvement involves enhancing the internal consistency of KS4 images by running \texttt{SCAMP} on multiple images from the same field across different epochs and filters, rather than on single-epoch, single-chip images. In this multi-image approach, we could further leverage the \texttt{SCAMP}-derived \texttt{ASTRIRMS} parameter to improve cross-band astrometric precision, which represents the internal RMSE of input dataset.

A major limitation in the current QA process is the absence of a recovery mechanism for images that fail the initial quality checks. At present, approximately 30\% of input images are excluded, a large amount of which are not fundamentally flawed but could be retained with rederived astrometric solutions. KS4 pipeline currently uses a fixed \texttt{ahead} file for each observatory, which acts as an initial guess for the astrometric solution. However, optimal solutions can vary with observing conditions and epochs. Therefore, a more adaptive strategy is required to select alternative \texttt{ahead} files based on metadata and re-applied automatically to reprocess problematic images.

To address local astrometric misalignments, we introduced a deviation map that quantifies spatial variations in astrometric accuracy. While this has proven effective in flagging problematic regions in a few images, its utility is limited in most cases. In fact, when the reference source density is low, the deviation map may overestimate local errors, leading to over-rejection of sources. Therefore, this diagnostic tool should be applied conservatively. For future data releases, we aim to implement more rigorous chip-level QA procedures to eliminate the need for applying deviation maps.

\subsection{Photometric Calibration Improvements} \label{sec:5.2.Photometric Calibration Improvements}

The current photometric calibration procedure consists of two steps: (1) chip-image-level ZP homogenization, and (2) stacked-image-level ZP homogenization. Although these steps are conceptually redundant, both are necessary in our framework. This is primarily because the KS4 dataset includes images acquired across multiple epochs and observatories, making it essential to correct inter-image photometric inconsistencies prior to stacking. However, due to the absence of PSF homogenization during stacking, spatial ZP non-uniformities remain in the stacked images. To mitigate these residuals, we apply a secondary correction using a local ZP deviation map, which enables spatially dependent flux rescaling.

In most cases, the final photometric calibration achieves high precision. When compared to Gaia XP photometry, the RMSEs are approximately 0.03 mag across all bands down to a depth of 19~AB mag, with typical 5$\sigma$ limiting magnitudes reaching 22–23~AB mag in the $BVRI$ filters.

Nonetheless, several inconsistencies in the chip-level and stacked-image-level calibration could have introduced systematic errors that propagate into the final photometry. Currently, chip-level ZPs are derived using magnitude transformations based on the APASS and SMSS catalogs. Given that Gaia XP offers deeper magnitude coverage and uniform filter throughput, this approach is now considered suboptimal. To maintain consistency with subsequent calibration stages, it is preferable to adopt synthetic photometry from Gaia XP at the initial stage of the pipeline.

Furthermore, the current chip-level calibration does not utilize a bad pixel mask during photometric cleaning. In future reductions, bad pixel masks will be generated prior to the ZP homogenization step to prevent contamination from spurious detections. Another inconsistency lies in the choice of photometric aperture: chip-level ZPs are calibrated using \texttt{MAG\_AUTO}, while final calibrations adopt fixed 5 arcsec apertures. Although \texttt{MAG\_AUTO} has the advantage of estimating total flux independent of PSF variation, its large aperture leads to severe blending issues in stellar-dense regions, especially in the $R$- and $I$-bands. To address this, we plan to either reduce the Kron radius in \texttt{MAG\_AUTO} or transition to using fixed-aperture photometry from the beginning of the calibration.

\begin{figure}
  \centering
  \includegraphics[width=\columnwidth]{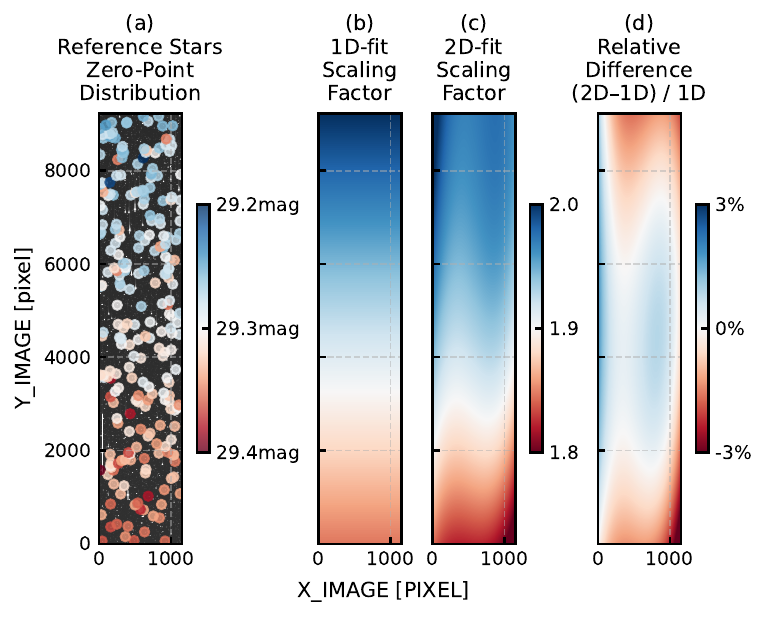}
  \caption{Comparison of ZP calibration across a single readout port.  
  \textbf{(a)} Distribution of ZPs measured from Gaia XP reference stars (colored by $\Delta m = m_{\rm Gaia}-m_{\rm KMTNet}$), overlaid on the image region.  
  \textbf{(b)} One-dimensional fit of the multiplicative ZP scaling factor as a function of detector $Y$ position.  
  \textbf{(c)} Two-dimensional fit of the ZP scaling factor across the focal plane.  
  \textbf{(d)} Residual map showing the fractional difference between the two fitted scaling factor maps ((2D–1D)/1D), highlighting local deviations up to $\pm3\%$.}
  \label{fig:12.zpcomp}
\end{figure}

An additional area for improvement lies in the ZP fitting methodology. The current chip-level ZP homogenization employs a one-dimensional linear fit along the y-axis, applied per readout port image. While this approach generally yields stable results, it is insensitive to photometric gradients in other directions. To better capture two-dimensional spatial variations in the ZP across the detector, we tested a two-dimensional polynomial fitting scheme. The results, presented in Figure~\ref{fig:12.zpcomp}, show that the resulting variations in flux scaling factors are typically minor.

Implementing the improvements described above is expected to reduce the number of frames excluded due to failed chip-level photometric homogenization and improve the overall consistency and robustness of the calibration process. Notably, Figure~\ref{fig:8.overallstats} shows that approximately 3\% of $I$-band images exhibit a photometric RMSE exceeding 0.07 mag. The calibration accuracy in these stellar-dense regions requires further investigation and validation.

\subsection{Stacked Image Processing Improvements} \label{sec:5.3.Stacked Image Processing Improvements}

While the current stacking procedure produces science-ready reference images, there is room for further refinement to improve both photometric uniformity and artifact suppression. Currently, the KS4 pipeline employs a simple median combination scheme for stacking, which is robust to outliers but does not fully exploit the available per-pixel quality information. Transitioning to a weighted mean combination using appropriately generated bad pixel masks and inverse-variance maps could lead to improved photometric residuals, especially in areas affected by non-uniform background noise or detector defects.

Beyond stacking methods, the selection criteria for individual frames can be further optimized. In the current release, we utilized a conservative FWHM threshold of 6 arcsec in the initial quality check (Section~\ref{sec:2.2.KMTNet Data Acquisition and Preprocessing}) and stacked all image sets that successfully passed the astrometric QA and photometric ZP scaling processes. While the resulting median seeing of KS4 DR1 is approximately 2 arcsec, implementing a stricter selection (e.g., FWHM $<$ 2 arcsec) in future releases will enhance the effective resolution and reliability of the KS4 reference images by replacing lower-quality frames with data obtained under superior conditions.

In addition, the uncertainty quantification of the stacked images can be further improved. Although the \texttt{SWarp} stacking software produces weight maps alongside the stacked images, these are not yet utilized within the current KS4 pipeline. These maps encode pixel-level reliability and could be instrumental in constructing quality masks or spatially resolved uncertainty maps. Incorporating such information would enable more precise error estimation across the stacked image, especially in regions affected by cosmetic defects or incomplete data.

Contaminant masking using the bad pixel map also requires improvements. In the current implementation, regions flagged in a master mask image are interpolated over on the stacked image using surrounding pixel values. While this approach is straightforward, it often introduces noticeable interpolation artifacts. A more effective strategy is to perform bad pixel correction at the single-exposure level prior to stacking. When the cleaned frames are subsequently combined, dithering naturally smooths out residuals from interpolation, leading to a cleaner final product.

Full-region interpolation of contaminating sources, such as bleeding trails, is suboptimal for preserving image reliability. If the bleeding pattern can be accurately modeled, their impact could be directly subtracted at the pixel level with minimal residuals. This principle also applies to cross-talk artifacts. Although the current cross-talk correction scheme is applied uniformly, significant residuals persist in some images. Future improvements should focus on deriving and applying image-specific cross-talk coefficients to minimize these artifacts. Enhancing the contaminant removal process in this way would elevate the quality and scientific utility of the KS4 reference images.

\begin{figure*}[!htbp]
\centering
\includegraphics[width=\textwidth]{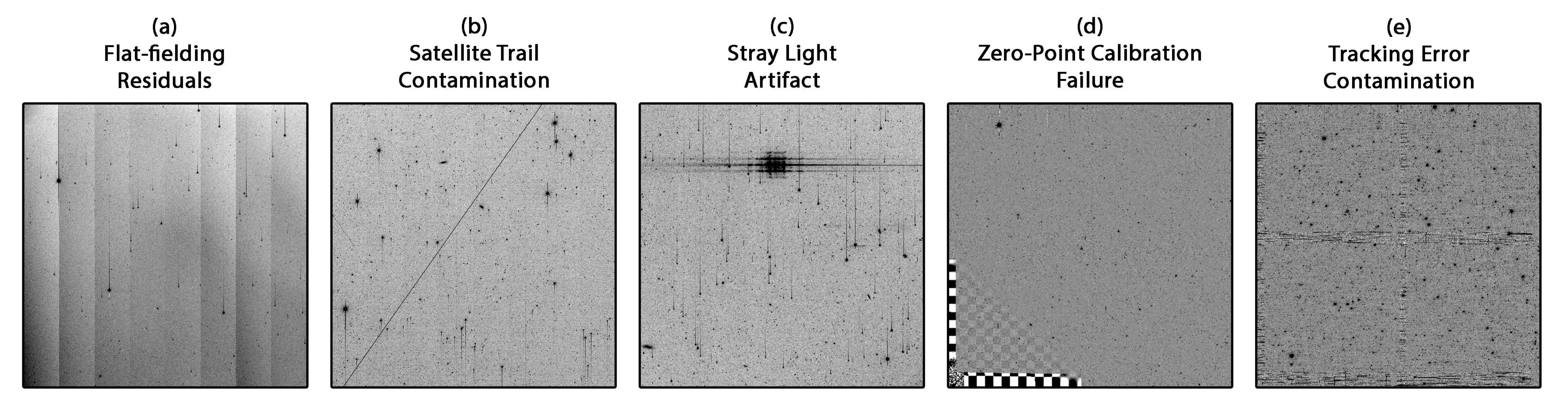}
\caption{Examples of image defects commonly encountered in KMTNet data, shown from left to right. \textbf{(a)} Flat-fielding residuals are caused by poor flat calibration under non-photometric conditions and often appear as distinctive vertical stripe patterns across the image. \textbf{(b)} Satellite trail contamination appears as bright linear streaks caused by satellites crossing the field during exposure. \textbf{(c)} Stray light artifacts result from internal reflections of bright out-of-field stars, introducing diffuse or structured illumination patterns. \textbf{(d)} Zero-point calibration failure arises when incorrect photometric scaling produces abnormally high or low pixel values in certain readout ports, leading to checkerboard patterns or strongly suppressed flux regions in the stacked image. \textbf{(e)} Tracking error contamination caused by a single exposure affected by tracking drift, resulting in elongated sources. The effect is most apparent in CCD gap regions where dithering coverage is minimal. Note that panels \textbf{(a)}--\textbf{(c)} show single-chip images, whereas panel \textbf{(d)}--\textbf{(e)} show the full stacked KS4 field.}
\label{fig:13.issues}
\end{figure*}

\subsection{Known Issues} \label{sec:5.4.Known Issues}

Despite the calibration procedures described above, several unresolved issues remain that can significantly influence source detection and photometry. These residual effects arise from instrumental artifacts or imperfect calibration processes that are challenging to fully characterize or correct. To aid users in identifying affected regions, we conducted visual inspections to flag problematic fields and annotated them accordingly. The list of flagged fields is available in the source code repository\footnote{\url{https://github.com/jmk5040/KMTNet_ToO/blob/main/config/KS4_DR1_Field_Issue_Flags.csv}}.

\subsubsection{Flat-Fielding Issue} \label{sec:5.4.1.Flat-Fielding Issue}

In some cases, exposures obtained under poor weather conditions exhibit irregular background structures due to imperfect flat-field calibration. These residual patterns, often caused by time-variable atmospheric conditions or flat-field inconsistencies, result in non-uniform background levels that can bias local photometric measurements.

The appropriate approach to handling such images is to identify and exclude them as poor-quality data. Our tests indicate that background fluctuations between readout ports serve as a reliable indicator of flat-fielding errors. However, this automated filtering process was not implemented in DR1. 

In this release, the visually inspected quality flag provides the most reliable criterion for mitigating the impact of this issue. We identified 32 out of 979 KS4 fields as affected, with the effect being more pronounced in longer-wavelength bands. Photometric measurements in these fields should therefore be interpreted with caution, as background irregularities may introduce systematic biases.

\subsubsection{Satellite Trail Contamination} \label{sec:5.4.2.Satellite Trail Contamination}

Satellite trails frequently leave residual streaks in stacked images, particularly in areas with low dithering redundancy. This issue is especially pronounced in CCD gap regions that are covered by only one or two exposures, where artifact rejection is limited due to the use of a median combination method. These imprints can result in spurious detections or biased flux measurements.

To mitigate such effects in KS4 DR1, we applied a catalog-level filtering procedure that identifies and flags spurious sources caused by residual satellite trails. This method detects regions with abnormally high densities of unmatched sources relative to reference catalogs and flags them as spurious. The detailed implementation and performance of this removal process are described in Section 3.2.1 of \citet{2026JKAS...Chang}.

A more robust approach involves detecting satellite trails at the single-exposure level and incorporating them into the bad pixel masks prior to stacking. Several algorithms for satellite trail detection have been developed and are already in use in major survey pipelines (e.g., \citealt{2024A&A...692A.199S}). Integrating such methods into the KS4 pipeline would significantly improve the quality of the stacked images by reducing false positives and photometric artifacts. 

\subsubsection{Stray Light Contamination} \label{sec:5.4.3.Stray Light Contamination}

Stray light contamination, presumed to originate from bright stars inside or outside the FOV, can introduce diffuse and structured illumination across the image. This contamination alters the local background, especially near the edges of the frame, and may degrade both photometric precision and source detection completeness in affected regions. Among the various image defects, stray light is particularly severe and ubiquitous, making it one of the primary unresolved limitations in achieving clean and reliable source detection.

Although the exact mechanism behind stray light is not fully understood, we have observed several consistent behaviors that provide some clues. First, the position of stray light shifts across exposures taken with different dithering patterns, implying that the effect is closely tied to the telescope optics or internal scattering paths. Second, fields affected by stray light often have bright stars located nearby, frequently aligned in right ascension or declination with the direction of the artifact. This spatial coincidence may suggest that those bright stars are the source of the scattered light.

Despite these indicators, the behavior of stray light remains complex, varying in both intensity and morphology. Although 19 KS4 DR1 fields have been visually identified and flagged as affected, additional contaminated fields likely remain undetected. Further improvements to the reduction pipeline will focus on automating the detection and masking of stray-light-affected regions to more effectively reduce photometric biases and spurious detections.

\subsubsection{Zero-Point Calibration Failure} \label{sec:5.4.4.Zero-Point Calibration Failure}

This type of error arises from catastrophic failures in the ZP scaling process described in Section~\ref{sec:3.2.2.Zero-point Homogenization for Chip Images}. Although the pipeline was designed to exclude such failures after the process, two unfiltered error images were identified through visual inspection, affecting the final stacked images. Specifically, the KS4 fields \texttt{0537} and \texttt{0698} in the $R$-band exhibit this issue, with the affected regions showing significantly elevated photometric uncertainties. Photometric measurements in these fields should therefore be interpreted with caution.

We attribute this issue to the unintended use of intermediate products that were not replaced with their correctly calibrated versions. In these affected chip images, erroneous scaling factors caused a subset of readout ports to produce extreme pixel values, resulting in localized regions of abnormally high or low intensity. During the stacking process, these anomalies resulted in failed convolution and left conspicuous checkerboard-like artifacts as shown in Figure~\ref{fig:13.issues} (d).

\subsubsection{Tracking Error Contamination} \label{sec:5.4.4.Tracking Error Contamination}

During visual inspection, we identified that KS4 \texttt{2327} $R-$band stacked image was affected by a tracking error. In this case, the image was constructed from three normal exposures and one exposure exhibiting elongated source profiles. As a result, sources in the stacked image display noticeable tails, appearing more prominent in the CCD gap regions where less dithering occurred.

As described in Section~\ref{sec:2.2.KMTNet Data Acquisition and Preprocessing}, our pipeline includes a quality control procedure that rejects images with tracking errors based on the average \texttt{ELONGATION} value. For the problematic exposure, this value was measured to be 3.4, well above our exclusion threshold of 1.9. Thus, under the current pipeline, this frame would have been correctly rejected. However, this stacked image was generated in an earlier version of the pipeline and was not subsequently replaced by the corrected product, allowing the affected frame to persist into KS4 DR1. This field is flagged in the KS4 field issue table, and users are advised to exercise caution when using this field.

\section{Conclusion} \label{sec:6.conclusion}

In this paper, we present the development and validation of the KS4 data reduction pipeline, which addresses a long-standing need within the KMTNet community: the absence of a large-scale, homogeneous, and publicly available reference image database for time-domain astronomy. As a result of this effort, KS4 DR1 provides astrometrically and photometrically uniform $BVRI$ reference images covering approximately 4,000~deg$^2$ of the southern sky.

The KS4 pipeline performs a complete sequence of data reduction steps, including image quality control, astrometric and photometric calibration, bad pixel masking, image stacking, and DIA preparation. It delivers uniformly calibrated images with sub-arcsecond astrometric precision (median $\leq$0.4~arcsec) and 5$\sigma$ depths of $B = 22.7$, $V = 22.6$, $R = 22.8$, and $I = 22.1$~AB mag. These results demonstrate that the pipeline provides science-quality reference images suitable for ToO observations across the southern sky.

The scientific utility of KS4 has been demonstrated through multiple GW follow-up campaigns. The KS4 DR1 pipeline delivers a uniform set of deep reference images and incorporates an automated framework for real-time transient discovery. It continuously monitors newly uploaded KMTNet data, performs image reduction, and executes DIA to identify transient candidates in near real time. Leveraging these capabilities, we established a complete image-subtraction and vetting workflow that has been applied to significant GW events occurred during the LVK O4 run. Although no definitive kilonova signals were detected, the pipeline successfully identified genuine transient-like sources within the localization areas of the five GW events, demonstrating the operational readiness and scientific potential of KMTNet for time-domain astronomy.

Although KS4 DR1 delivers uniform and scientifically robust reference images,  several improvements are planned for future data releases. These include transitioning to Gaia XP synthetic photometry at earlier processing stages, adopting two-dimensional spatial models for ZP homogenization, and incorporating stacking weight maps to produce spatially resolved error maps and improved photometric uncertainty estimates. Artifact rejection, bad pixel handling, and contaminating source cleaning will also be improved through single-image level preprocessing.

The KS4 pipeline presented in this study is publicly available and can be applied to all KMTNet images with user-defined field configurations. The pipeline will be continuously updated to address the issues discussed above, and future data releases will expand the survey to encompass a larger data volume with enhanced calibration and quality assessment. KS4 thus establishes a robust foundation for a reference image library that will continue to facilitate fast and reliable transient discovery across the southern sky.

\acknowledgments

This work was supported by the National Research Foundation of Korea (NRF) grant No. 2021M3F7A1084525, funded by the Korea government (MSIT). BP acknowledges support from the NRF grant (RS-2025-00573214) funded by the MSIT. SWC acknowledges support from the Basic Science Research Program through the NRF funded by the Ministry of Education (RS-2023-00245013). G.S.H.P. acknowledges support from the Pan-STARRS project, which is a project of the Institute for Astronomy of the University of Hawai'i, and is supported by the NASA SSO Near Earth Observation Program under grants 80NSSC18K0971, NNX14AM74G, NNX12AR65G, NNX13AQ47G, NNX08AR22G, 80NSSC21K1572, and by the State of Hawai'i.

This research has made use of the KMTNet system operated by the Korea Astronomy and Space Science Institute (KASI) at three host sites of CTIO in Chile, SAAO in South Africa, and SSO in Australia. Data transfer from the host site to KASI was supported by the Korea Research Environment Open NETwork (KREONET).

This work has also made use of data from the European Space Agency (ESA) mission \textit{Gaia} (\url{https://www.cosmos.esa.int/gaia}), processed by the \textit{Gaia} Data Processing and Analysis Consortium (DPAC, \url{https://www.cosmos.esa.int/web/gaia/dpac/consortium}). Funding for the DPAC has been provided by national institutions, in particular those participating in the \textit{Gaia} Multilateral Agreement. 

The national facility capability for SkyMapper has been funded through ARC LIEF grant LE130100104 from the Australian Research Council, awarded to the University of Sydney, the Australian National University, Swinburne University of Technology, the University of Queensland, the University of Western Australia, the University of Melbourne, Curtin University of Technology, Monash University and the Australian Astronomical Observatory. SkyMapper is owned and operated by The Australian National University's Research School of Astronomy and Astrophysics. The survey data were processed and provided by the SkyMapper Team at ANU. The SkyMapper node of the All-Sky Virtual Observatory (ASVO) is hosted at the National Computational Infrastructure (NCI). Development and support of the SkyMapper node of the ASVO has been funded in part by Astronomy Australia Limited (AAL) and the Australian Government through the Commonwealth's Education Investment Fund (EIF) and National Collaborative Research Infrastructure Strategy (NCRIS), particularly the National eResearch Collaboration Tools and Resources (NeCTAR) and the Australian National Data Service Projects (ANDS).

We express our sincere gratitude to the observers and technical staff of KMTNet for their continuous efforts in collecting the observational data that made this study possible. We gratefully acknowledge Dr. Seung-Lee Kim for insightful discussions on the origin of stray light and his assistance in tracing its likely sources. We also thank Dr. Sang-Mok Cha and Dr. Eon-Chang Sung for their valuable input on interpreting image artifacts and contributing to diagnostic improvements for image quality control.

\appendix

\section{Electronic Noise Pattern in SAAO Images} \label{sec:A.Electronic Noise Pattern}

\begin{figure*}
\centering
\includegraphics[width=\textwidth]{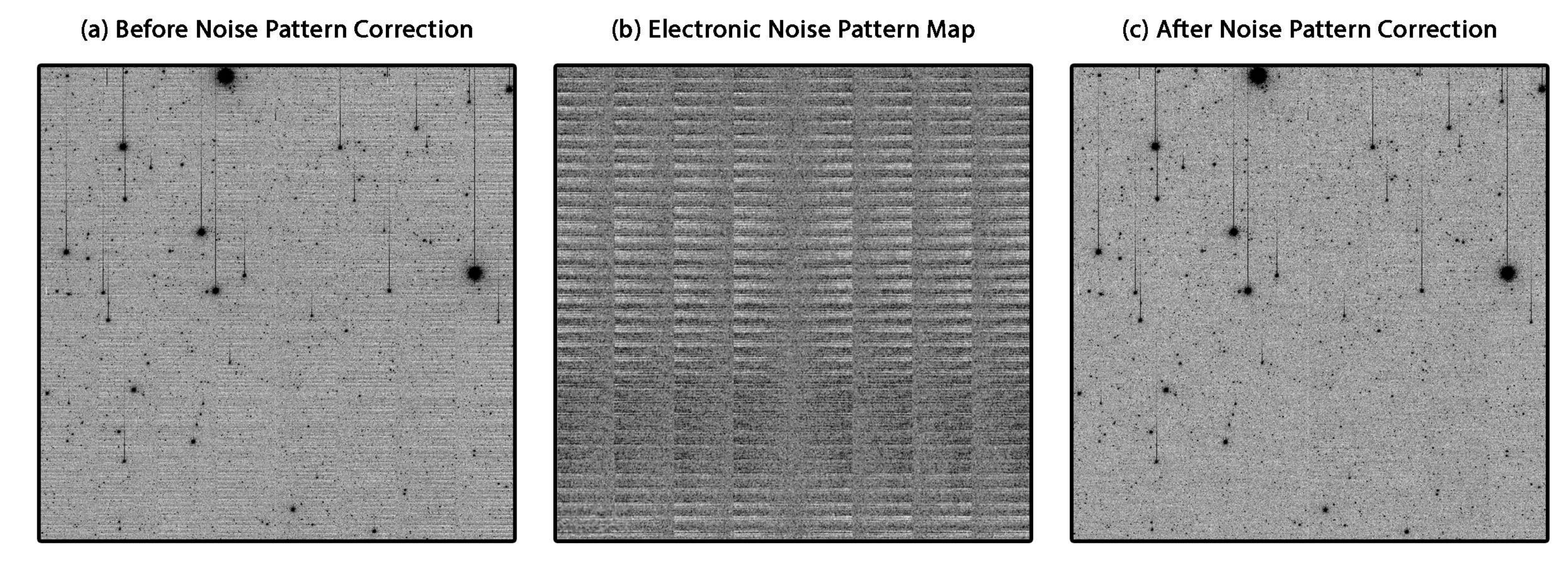}
\caption{Example of the electronic noise pattern identified in an SAAO chip image obtained on 2022 January 14.
\textbf{(a)} Raw image exhibiting a symmetric horizontal noise pattern across the eight readout ports.
\textbf{(b)} Master noise pattern map generated by median-combining residuals of the readout ports after masking all sources and subtracting the modeled background.
\textbf{(c)} Corrected image after subtracting the noise map, demonstrating substantial suppression of the patterned artifact.}
\label{fig:A1.SAAO_noise}
\end{figure*}

A horizontal electronic noise pattern was identified in a subset of SAAO images, particularly in data obtained during the 2021–2022 period. This artifact, likely caused by power-supply instability or readout interference, appears as a left-right (east–west) symmetric, repetitive horizontal structure across the eight readout ports of the CCD, centered about the chip midline (Figure~\ref{fig:A1.SAAO_noise}).

This electronic noise pattern has been systematically mitigated in the KS4 DR1 images. The correction procedure uses object-masked residual frames generated with the \texttt{CHECK\_IMAGE = -OBJECT} option in \texttt{SExtractor}. Each chip image is divided into eight segments corresponding to the readout-port boundaries. The first four segments are horizontally flipped to align with the symmetry of the latter four, and all eight segments are then median-combined to construct a master noise template. This template is tiled to match the full chip extent, producing a noise-pattern map for each SAAO image. 

Subtracting this map from the original image yields substantial suppression of the artifact, reducing the background standard deviation by 10–20\% in affected frames. Although faint residuals may persist in extreme cases, the correction provides a consistent improvement in the quality of SAAO data and is recommended when processing images from this site.

\bibliography{KS4DR1}

\end{document}